\documentclass[fleqn,usenatbib]{mnras}

\usepackage{newtxtext,newtxmath}
\usepackage[T1]{fontenc}

\DeclareRobustCommand{\VAN}[3]{#2}
\let\VANthebibliography\thebibliography
\def\thebibliography{\DeclareRobustCommand{\VAN}[3]{##3}\VANthebibliography}

\usepackage{graphicx}	% Including figure files
\usepackage{amsmath}	% Advanced maths commands
\usepackage[normalem]{ulem}

\title[The cosmic web and galaxy evolution]{The effect of cosmic web filaments on galaxy evolution}

\author[C. J. O'Kane et al.]{
Callum J. O'Kane$^{1}$\thanks{E-mail: callum.o'kane@nottingham.ac.uk},
Ulrike Kuchner$^{1}$,
Meghan E. Gray$^{1}$,
and Alfonso Arag\'{o}n-Salamanca$^{1}$
\\
$^{1}$School of Physics, Astronomy, University of Nottingham, Nottingham NG7 2RD, UK
}

\date{Accepted XXX. Received YYY; in original form ZZZ}

\pubyear{2024}

\begin{document}
\label{firstpage}
\pagerange{\pageref{firstpage}--\pageref{lastpage}}
\maketitle

\begin{abstract}
Galaxy properties are known to be affected by their environment. This is well established for the extremes of the density scales, between the high-density cluster environment and the low-density field. It is however not fully understood how the intermediate-density regime of cosmic web filaments affects galaxy evolution. We investigate this environmental effect using a mass complete sample of 23,441 galaxies in the Sloan Digital Sky Survey DR8 Main Galaxy Sample (${M}_{\text{Stellar}} > 10^{9.91} \text{M}_{\sun}$). We define 6 environments, probing different density regimes and representing unique stages in the structure formation process, comparing the differences in star formation activity and morphology between them. We find that galaxies in filaments tend to be less star forming and favour more early-type morphologies than those in the field. These differences persist when considering stellar mass-matched samples, suggesting that this is a consequence of the environment. We further investigate whether these trends are a result of the large scale or local environment through constructing samples matched both in stellar mass and local galaxy density. We find that when also matching in local galaxy density, the differences observed between the filament and field population vanishes, concluding that the environmental effect of filaments can be entirely parameterised by a local galaxy density index. We find that differences can still be seen in comparisons with the interiors of clusters, suggesting these are unique environments which can impart additional physical processes not characterised by local galaxy density.

\end{abstract}

% Select between one and six entries from the list of approved keywords.
% Don't make up new ones.

%%%%%%%%%%%%%%%%%%%%%%%%%%%%%%%%%%%%%%%%%%%%%%%%%%

%%%%%%%%%%%%%%%%% BODY OF PAPER %%%%%%%%%%%%%%%%%%

\section{Introduction}

Under a cold dark matter cosmological paradigm, the present-day matter distribution is the result of the compounding clustering of dark matter halos. This is seen in galaxy clusters, maxima of the universal density field and the culmination of the hierarchical formation model. These clusters, and the elongated chains of galaxies called `filaments' which connect them, form a highly complex and anisotropic matter distribution at the megaparsec scale called the Cosmic Web \citep{bond_how_1996}. The Cosmic Web, described by Zel'Dovich formalism \citep{zeldovich_gravitational_1970}, comprises of clusters, filaments, sheets, walls, knots, and voids, each representing unique stages of the structure formation process. This large-scale structure (LSS) of the universe is well established in both theory and observations, with observational signatures detected in many large-scale surveys such as the Sloan Digital Sky Survey \citep[SDSS;][]{york_sloan_2000}, GAMA \citep{driver_gama_2009} and 2dFGS \citep{colless_2df_2001}. Filaments of the cosmic web account for a large portion of the universal mass budget, with approximately half of the mass located within filaments despite occupying only 6\% of the total volume \citep{cautun_evolution_2014}. The role of this large scale structure in shaping galaxy properties is relatively unknown, and must be considered for a comprehensive theory of galaxy evolution.

The process in which a galaxy truncates its star formation, transitioning from actively star-forming to quiescent, is often called `quenching' and is not entirely understood. This is due to its complex nature, with numerous different processes often occurring simultaneously to halt star formation. These processes generally fall within two different categories, mass quenching and environmental quenching. Mass quenching refers to internal processes such as Active Galactic Nuclei feedback \citep{croton_many_2006} and Supernovae explosions \citep{larson_effects_1974}. Environmental quenching refers to the effect caused by the environment where the galaxy is located, including processes such as ram pressure stripping \citep{gunn_infall_1972}, strangulation \citep{larson_evolution_1980} and harassment \citep{moore_galaxy_1996}. It has been found that the effects of mass and the environment appear to be separable up to at least $z=1$, suggesting that both mass and environmental quenching are independent processes \citep{peng_mass_2010}.

\begin{table*}
    \centering
\caption{A description of the 3 regions considered in this work. 1. Name of the region used throughout this work. 2. Name of the WWFCS targets within the regions. 3. Redshift of the cluster. 4. Right ascension of the cluster (J2000). 5. Declination of the cluster (J2000). 6. [Min, Max] redshift of the region. 7. Right Ascension of region centre. 8. Declination of the region centre. 9. [Min, Max] box coordinates in angular coordinates $(\alpha-\alpha_{\text{centre}})\cos\delta$. 10. [Min, Max] box coordinates in angular coordinates ($\delta - \delta_{\text{centre}}$) 11. Number of galaxies in the region following the cuts detailed in \autoref{Sec:Data}.}
\label{tab: Field Description}
    \begin{tabular}{|l|l|c|c|l|l|c|c|c|l|l|c|} \hline  
            Region $^{\text{1}}$& &\multicolumn{2}{|c|}{$\textbf{A}$} && &\multicolumn{3}{|c|}{$\textbf{B}$} &&  &$\textbf{C}$\\ \hline  
            WWFCS target name$^{\text{2}}$& &Z2844&  RX1022 && &A1668&  A1795&  A1831 &&  &A2124\\ \hline  
            $z_{\text{cluster}}$$^{\text{3}}$& &0.050&  0.055 && &0.063&  0.063&  0.063 &&  &0.067\\ \hline 
 $\alpha_{\text{cluster}}$$^{\text{4}}$& & 150.65$\degr$& 155.54$\degr$ && & 195.94$\degr$& 207.21$\degr$& 209.81$\degr$ && &236.24$\degr$\\ \hline 
 $\delta_{\text{cluster}}$$^{\text{5}}$& & 32.71$\degr$& 38.53$\degr$ && & 19.27$\degr$& 26.59$\degr$& 27.98$\degr$ && &36.11$\degr$\\\hline \hline
    $z$$^{\text{6}}$& &\multicolumn{2}{|c|}{[0.0425, 0.0625]} && &\multicolumn{3}{|c|}{[0.053, 0.073]} && &[0.057, 0.077]\\\hline
 $\alpha_{\text{centre}}$$^{\text{7}}$& & \multicolumn{2}{|c|}{153.09$\degr$} && & \multicolumn{3}{|c|}{202.88$\degr$} && &236.24$\degr$\\\hline
 $\delta_{\text{centre}}$$^{\text{8}}$& & \multicolumn{2}{|c|}{35.61$\degr$} && & \multicolumn{3}{|c|}{23.62$\degr$} && &36.11$\degr$\\ \hline 
 $(\alpha-\alpha_{\text{centre}} )\cos\delta$$^{\text{9}}$& & \multicolumn{2}{|c|}{[$-16.376\degr$, $16.376\degr$]}& & & \multicolumn{3}{|c|}{[$-18.207\degr$, $18.207\degr$]}& & &[$-10.706\degr$, $10.706\degr$]\\ \hline 
 $(\delta-\delta_{\text{centre}})$$^{\text{10}}$& & \multicolumn{2}{|c|}{[$-16.964\degr$, $16.964\degr$]}& & & \multicolumn{3}{|c|}{[$-15.693\degr$, $15.693\degr$]}& & &[$-10.706\degr$, $10.706\degr$]\\\hline
 $N^{\text{11}}$& & \multicolumn{2}{|c|}{6201} && & \multicolumn{3}{|c|}{11547} && &5693\\\hline 
    \end{tabular}
        
\end{table*}

The correlation between galaxy properties and local galaxy density is well established in the literature. One such example is the Morphology--Density relation \citep{dressler_galaxy_1980}, showing that galaxies with early-type morphologies are more abundant in regions of high local galaxy density, such as cluster cores. It is currently unclear as to whether relationships such as the Morphology--Density and others like it such as Colour--Density (e.g., \citealt{baldry_galaxy_2006, bamford_galaxy_2009}) and Star Formation--Density (e.g., \citealt{hashimoto_influence_1998,kauffmann_environmental_2004}) are driven entirely by processes correlated with local density on small-scales or are influenced by the geometry, topology, and physics of the LSS. The multiscale characteristics of the cosmic web means that its components span density scales over many magnitudes \citep{aragon-calvo_multiscale_2010}. This means that the different cosmic web components, including cosmic filaments, cannot be defined or identified through density alone.

Investigations concerning the effect of cosmic web filaments on galaxy properties are gaining momentum. Within the hierarchical formation model, the assembly histories of galaxies are expected to be affected by the past large-scale environment, with intrinsic properties, such as mass, spin and alignment affected by this history whilst also correlating with the present environment. Although it is generally agreed that filaments do affect the properties of galaxies to some degree, the mechanisms responsible and their relative contributions are not well known. Many studies find that galaxies closer to filaments are redder in colour with reduced star formation (e.g., \citealt{chen_detecting_2017,kraljic_galaxy_2018,martinez_galaxies_2016,singh_study_2020,mahajan_ultraviolet_2018,Castignani_Virgo_1, parente_star_2023}). For example, \citet{kuutma_voids_2017} find an elevated Elliptical-to-Spiral ratio towards filaments, concluding that this may be evidence of an increased merger rate inside filaments, transforming spiral galaxies to ellipticals as they migrate towards clusters. Possibly related to this, \citet{chen_detecting_2017} find that close to and inside filaments, galaxies are larger than further away from filaments. 

Filaments host large reservoirs of multiphase gas with varying temperatures and densities \citep{snedden_star_2016}. This extra gas component could affect galaxies in a unique way which cannot be characterised through local galaxy density alone. One such example is shown by \citet{kleiner_evidence_2017}, who find that the most massive galaxies ($\log ({M}_{\text{stellar}}/{M}_{\sun}) > 11 $) possess enhanced H{\sevensize I} fractions relative to the field population, suggesting that sufficiently massive galaxies can rejuvenate their gas supply through accretion from filaments, an example of `Cosmic Web Enhancement' \citep{vulcani_gasp_2019}. Supporting this, other studies report that galaxies in filaments have enhanced rates of star formation (e.g., \citealt{darvish_cosmic_2014,fadda_starburst_2008}), although sample sizes, cosmic variance, and different characterisations of the environment certainly make it hard to settle on a conclusion. Furthermore, filamentary cold gas accretion is especially relevant at high redshift during galaxy formation. Studies at higher redshifts such as that of \cite{darvish_cosmic_2014}, investigating filamentary structures at $z=0.84$, find that, while the median mass and star formation rate of individual star-forming galaxies do not depend on the environment, the fraction of star-forming galaxies is elevated in filaments. The authors propose that mild galaxy-galaxy interactions may be responsible or that this enhancement could be the result of selection biases.  

In addition to filamentary accretion, galaxy clusters accumulate a large fraction of their mass through the accretion of galaxy groups \citep{mcgee_accretion_2009,dressler_imacs_2013}; these groups assemble inside filaments and drift towards clusters. During their accretion into cluster cores, galaxies are affected by the cluster's hot intra-cluster gas (the intra-cluster medium, ICM). While it seems evident that galaxies undergo some transformations (that changes galaxies from star-forming late-type galaxy to passive early-type galaxy) as they interact with the ICM (e.g. \citealt{gunn_infall_1972,nulsen_transport_1982}), a transformation could, at least in part, occur before the galaxies reach the cluster. This idea is called `pre-processing' \citep{fujita_pre-processing_2004}, and has gained considerable attention over the past few decades. Two possible environments in which pre-processing may occur are filaments and groups. For example, \citet{donnan_role_2022} report an increased gas-phase metallicity for galaxies closer to nodes of the cosmic web relative to those further away, with a similar, weaker trend observed for filaments. \citet{martinez_galaxies_2016} provides evidence for the scenario that both web components relate to pre-processing. Comparing galaxies in filaments with galaxies undergoing isotropic infall onto clusters, they found that filaments contain a larger fraction of galaxies with reduced specific star formation rates than those isotropically infalling. 

Simulation work has also been employed to understand the impact of filaments better. One such example is \citet{bulichi_how_2023} using the Simba simulations to find that at $z=0$ galaxies within 100 kpc of filaments are significantly suppressed in star formation, a similar result is also seen in EAGLE, and IllustrisTNG simulations. The authors conclude that this may be the result of shock-heating within filaments. This is further shown by \citet{hasan_how_2023}, who use IllustrisTNG simulations and find that at $z \leq 0.5$ low-mass galaxies ($8 \leq \log({M}_{\text{stellar}}/{M}_{\sun}) < 9$) are significantly suppressed in star formation within 1 Mpc to filaments, a trend driven mostly by satellite galaxies. This provides further evidence that low-mass galaxies may be more susceptible to environmental quenching compared to higher-mass galaxies.

A key unanswered question that our paper addresses, is whether or not the effects of filaments are solely a consequence of the well established relations with local galaxy density (e.g. Morphology--Density and Star Formation--Density) or if the physical processes associated with the large-scale cosmic web imparts specific effects. The scale over which one considers densities correlates with different processes; at the smallest scales, density corresponds with the most stochastic and recent processes, whereas larger scales consider the averaged, smooth histories of galaxies \citep{kraljic_galaxy_2018}. Past studies have attempted to take into account local density through various means, such as number density \citep{kraljic_galaxy_2018,eardley_galaxy_2015}, r-band luminosity density \citep{kuutma_voids_2017} and the Delaunay Tessellation Field Estimator \citep{galarraga-espinosa_flows_2023, laigle_cosmos2015_2018}. It is however unclear if galaxy density on small ($\leq$ Mpc) scales can explain the observed trends within filaments. If galaxies in filaments are subject to processes which are not characterised by local galaxy density, signatures of this could manifest as differential effects at constant mass and density.

\begin{figure}
	\includegraphics[width=\columnwidth,keepaspectratio]{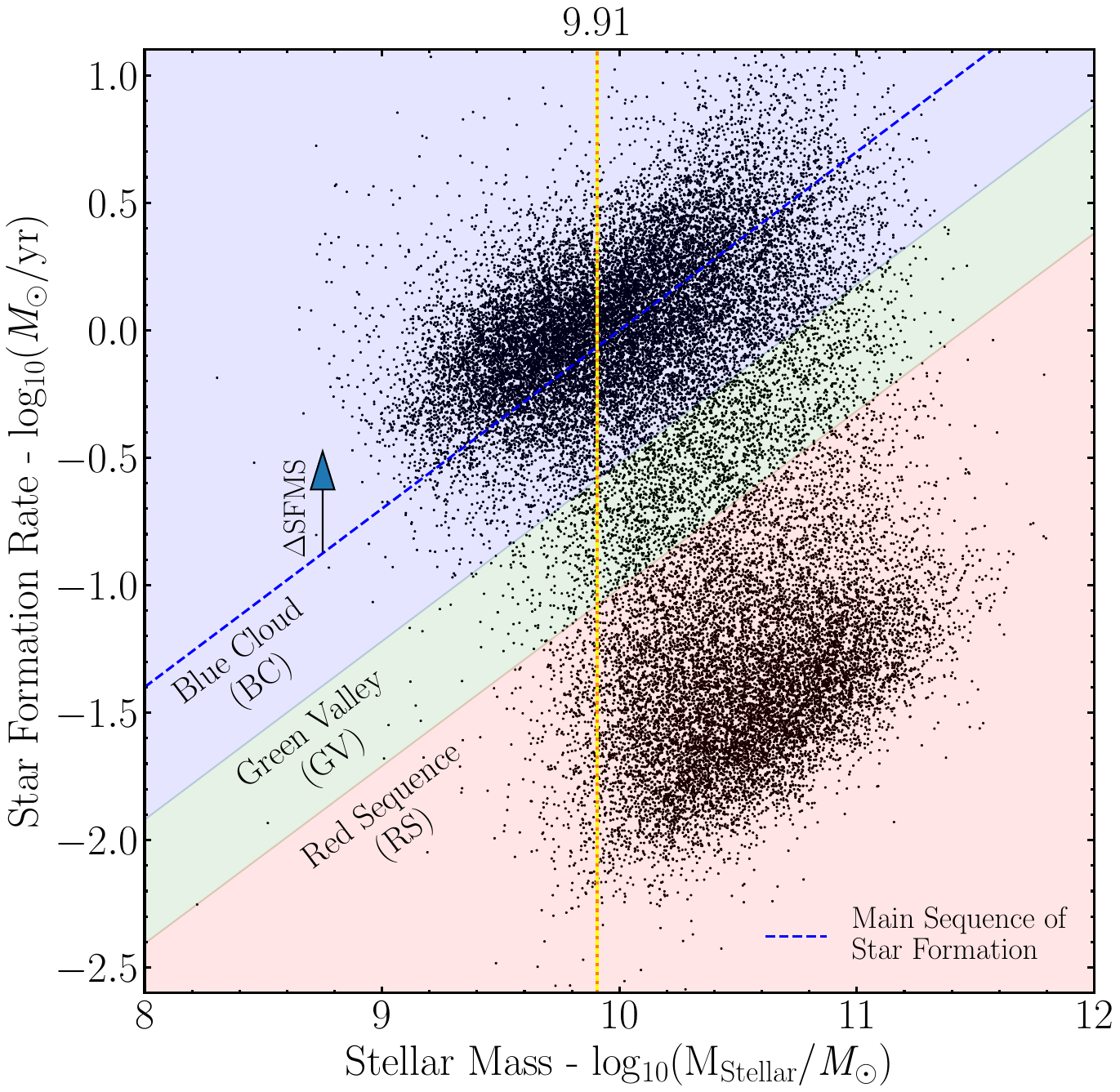}
    \caption{Star formation rate as a function of stellar mass for the galaxies considered in this work. We denote the 90\% completeness limit at $z = 0.077$ with the yellow vertical dashed line. Shaded in blue is the Blue Cloud (BC), in green is the Green Valley (GV), and in red is the Red Sequence (RS). The dashed blue line denotes the Main Sequence of star-forming galaxies, calculated using \autoref{MS Equation} with $b = -7$. An illustrative arrow describes the metric of Star Formation Suppression $\Delta$SFMS, used throughout this work. See text for details.}
    \label{fig:SFRvsMass}
\end{figure}

Observational studies are limited by the available data. Current spectroscopic surveys either lack the necessary depth or spatial extent for a robust investigation of pre-processing in the vicinity of galaxy clusters. As such, the majority of past studies are restricted to large-scale surveys (e.g., SDSS). While they offer large statistical samples, they generally suffer from relatively low sampling rates. Conversely, higher-sampling in dedicated studies investigating pre-processing of individual clusters are currently restricted to only the most nearby ones, such as the Virgo cluster (e.g., \citealt{Castignani_Virgo_1,Castignani_Virgo_2,brown_vertico_2023,chung_star-forming_2021}). To address these requirements, we must look ahead to the next generation of spectroscopic surveys. One such survey of note is the WEAVE Wide-Field Cluster Survey (WWFCS; Kuchner et al., in prep.) utilising the WEAVE (WHT Enhanced Area Velocity Explorer) multi-object spectrograph on the William Herschel Telescope \citep{jin_wide-field_2023, Dalton_WEAVE}. The WWFCS will systematically observe 16--20 galaxy clusters in the nearby universe $(z\sim0.05)$ out to $5R_{200}$, measuring thousands of optical spectra per cluster down to a total magnitude limit of $r \approx 19.75$, corresponding to a stellar mass limit $\log({M}_{\text{stellar}}/{M}_{\sun}) \approx 9$. The large spatial coverage with deep observations will prove invaluable in studies of the environmental effects of the cosmic web around galaxy clusters. 

This paper serves as a preliminary investigation to the WWFCS, taking a step towards answering the question of local density vs large scale structure in the context of environmental quenching, with the currently available data. We explore the effect of cosmic web filaments on galaxy properties in the Sloan Digital Sky Survey DR8 Main Galaxy Sample, using \texttt{DisPerSE} \citep{sousbie_persistent_2011} to map the projected filament network. We anchor our investigation around 6 of the target clusters planned to be observed in the WWFCS with sufficient coverage by SDSS. The SDSS is ideal for such studies as it is thoroughly complete at this redshift range whilst spanning a large angular size of 7966 deg$^2$. Although it doesn't reach the depth and sampling density of the planed WWFCS, it allows us to cover very large areas with a homogeneous dataset to explore a broad range of environments. We define 6 distinct environments and test differences in star formation activity and morphology, investigating how these differences vary in samples matched in stellar mass only, and samples matched both in stellar mass and local galaxy density.

This paper is outlined as follows. Section~\ref{Sec:Data} outlines the data and the sample selection used in this work. Section~\ref{Sec: Methodology} presents the main methodology, including the mapping of filaments, the star formation suppression metric that we use to characterise star formation activity, the definition of local densities, and the environmental classification scheme. Results are presented and discussed in Section~\ref{sec: Results}, with conclusions and future extensions presented in Section~\ref{sec: Conclusions}.

Throughout this work, we adopt the \textit{WMAP 9} cosmology \citep{hinshaw_nine-year_2012}, with ${H}_0 = 69.32\,$km$\,$s$^{-1}\,$Mpc$^{-1}$. All statistical errors are determined via bootstrapping, such that the error corresponds to the standard deviation of the distribution of the target quantity calculated in 1000 sub-samples of the parent, whilst allowing for replacement.

\section{Data} \label{Sec:Data} 

In this work, we select galaxies from the Sloan Digital Sky Survey Data Release Eight (SDSS DR8, \citealt{aihara_eighth_2011}) belonging to the Main Galaxy Sample\footnote{While the Main Galaxy Sample was first observed and released in DR7 and the sample is practically unchanged in DR8, we query galaxies from DR8 to utilise the reprocessed photometry as well as the availability to retrieve the MPA--JHU spectral measurements (see section \ref{MassesandSFR}) from SDSS directly.}. We select galaxies from the Main Galaxy sample with \texttt{class}$\,=\,$``\texttt{GALAXY}'' and \texttt{zWarning}$\,= 0$ or 16, indicating reliable redshifts. We further limit our sample to galaxies with an extinction corrected apparent r-band petrosian magnitude $r_{\text{petro}}$ less than $17.77$, corresponding to the completeness limit of the SDSS Main Galaxy Sample \citep{strauss_spectroscopic_2002}.

\subsection{Galaxy Sample}

This investigation is motivated by the WWFCS, exploring what may be inferred from existing data over larger spatial scales, both to identify the current state of knowledge and as a measure to build upon. We therefore choose to anchor our investigation and galaxy sample around the WWFCS target clusters. These target clusters consists of galaxy clusters that have previously been observed in the WINGS \citep{fasano_wings_2006} and OmegaWINGS \citep{moretti_omegawings_2017} surveys. These WINGS clusters cover a wide range of velocity dispersions, X-ray luminosities, and virial masses  ($\sigma = 500$--$1200\,\text{km\,s}^{-1}$; $\log L_\text{X} = 43.3$--$45.0\,\text{erg\,s}^{-1}$; $\log_{10} (M/M_{\sun}) = 13.8$--$15.5$). The WWFCS cluster targets are selected such that they possess statistically indistinguishable velocity dispersion and X-ray luminosity distributions from the parent WINGS sample, forming an unbiased sample in terms of their mass distribution.

Of the 16 WWFCS targets, 12 reside within the SDSS Main Galaxy sample footprint. Given the reduced number density of SDSS galaxies compared to the number density which will be observed in the WWFCS, we must observe larger spatial scales to accommodate for the reduced statistics. Motivated by this, we initially select galaxies within a $100\times100\,$Mpc$^2$ area centred on the WWFCS targets, within a redshift range $\Delta z = \pm 0.01$ centered on the cluster redshift. Of the 12 targets, only 6 reside at least $50\,$Mpc away from the main galaxy sample footprint edges, allowing for the above selection to occur. Since there is some overlap between these 6 areas, to avoid including duplicate galaxies, we opt to merge the overlapping areas, resulting in 3 unique regions of different sizes with a depth of $\Delta z = 0.02$, labelled A, B and C. The resulting regions, as well as the WWFCS targets from which they are defined, are described in \autoref{tab: Field Description}. While these regions are defined according to the WWFCS cluster target locations, we emphasise that these WWFCS targets are not inherently unique, and they form an unbiased subsample of a complete parent cluster sample. Therefore, these WWFCS targets are not treated any differently to other groups and clusters introduced in the subsequent analysis and are used only in the  selection of the location of the galaxy sample we analyse.

We find that individual results of each region are comparable and statistically compatible with each other. Moving forward, We therefore opt to stack the galaxy samples of all three regions to enhance our overall statistics. This results in an initial sample of 32,975 galaxies.

\subsection{Group and cluster membership}

To classify galaxies as members of groups and clusters, we adopt the SDSS DR7 galaxy group catalogue of \citet[hereafter YGC]{yang_galaxy_2007}. This is a catalogue of groups produced from applying an iterative halo-based group finder to the New York University Value added Catalogue \citep{blanton_new_2005}. The YGC catalogue uses a friend-of-friends algorithm to identify tentative groups, and makes estimates of halo mass, size and velocity dispersion. These estimates are used to update the group membership,  and the estimates are then redetermined. The process is repeated until there is no further change in membership. In this work we use the halo mass estimate based on ranking group luminosity, as provided by the YGC. In what follows we define a simple cutoff, and call `groups' those systems from the YGC with halo masses $10^{13}\text{M}_{\sun} < M_{\text{h}} < 10^{14}\text{M}_{\sun}$, and `clusters' the systems with $M_{\text{h}} > 10^{14}\text{M}_{\sun}$.

In this work we consider only groups with central redshifts contained within the redshift range of each of the regions described in \autoref{tab: Field Description}.
In addition, we must also consider the effect of groups and clusters just outside these boundaries. This is because the redshift distributions of galaxies within groups and clusters are elongated along the line-of-sight due to the Fingers-of-God effect \citep{jackson_fingers_1972}. This elongation could bring many of the galaxies residing in groups and clusters whose redshifts are just outside the region boundaries into the redshift bounds. In other words, these galaxies likely have cosmological distances that place them outside the desired range but extend inside because of the Finger-of-God effect. We therefore choose to remove them to avoid introducing erroneous structures. To identify these galaxy interlopers, we select all groups/clusters whose redshift is at most $2z_\sigma$ outside the redshift bounds, where $z_\sigma$ is the velocity dispersion of the group/cluster in redshift units. We then assign galaxies to these groups/clusters if they reside within $R_{180}$ projected distance of a group centre or $2.5R_{180}$ projected distance of a cluster centre (we justify this choice in section \ref{Methods: Clusters and Groups}) and if they satisfy $|z-z_{\text{group/cluster}}| < 2z_{\sigma}$. These galaxies likely have cosmological redshifts outside the desired range but have observed redshifts inside due to the peculiar motion induced by their host group/cluster. These galaxies may be misclassified as field or filament galaxies, rather than members of groups or clusters, and must be removed from our sample. Of the 32,975 galaxies in our initial sample, this process removes 852 (2.6\%) galaxies, leaving 32,123 galaxies in our sample.

\subsection{Masses and Star Formation Rates \label{MassesandSFR}}

We use stellar masses and star formation rates provided by the MPA--JHU catalogue and are retrieved as the quantities \texttt{log\_tot\_p50} and \texttt{sfr\_tot\_p50} in the SDSS table \texttt{galSpecExtra}. Masses were estimated following the methodology outlined in \cite{kauffmann_stellar_2003}. Star formation rates were calculated in accordance with \cite{brinchmann_physical_2004} and aperture corrected as described by \cite{salim_uv_2007}. Within the regions considered here, of the total 32,123 galaxies, 100\% have stellar mass estimates and 32,016 (99.7\%) have star formation rate estimates. 

The SDSS Main Galaxy Sample is a magnitude-limited sample ($r_{\text{petro}} < 17.77$; \citealt{strauss_spectroscopic_2002}), leading to an increasing mass limit with redshift. To allow comparisons across redshifts, we construct a mass-limited sample. This is done by taking all galaxies within the main galaxy sample and constructing equi-populated bins in redshift containing 5000 galaxies. For each bin, we determine the 90$^{\rm th}$ percentile stellar mass. Applying a logarithmic fit of the form $\log_{10}(\text{M}_{\text{stellar}}/M_{\sun}) = A + B\log_{10}(z)$ to these percentiles over the redshift range considered here yields a 90\% completeness limit ${M}_{\text{stellar}} = 10^{9.91} \text{M}_{\sun}$ at $z = 0.077$, the largest redshift considered in this work. We retain only galaxies with a stellar mass exceeding this, leaving a total of 23,441 galaxies in our mass-limited sample.

\begin{figure*}
	\includegraphics[width=0.925\textwidth,keepaspectratio]{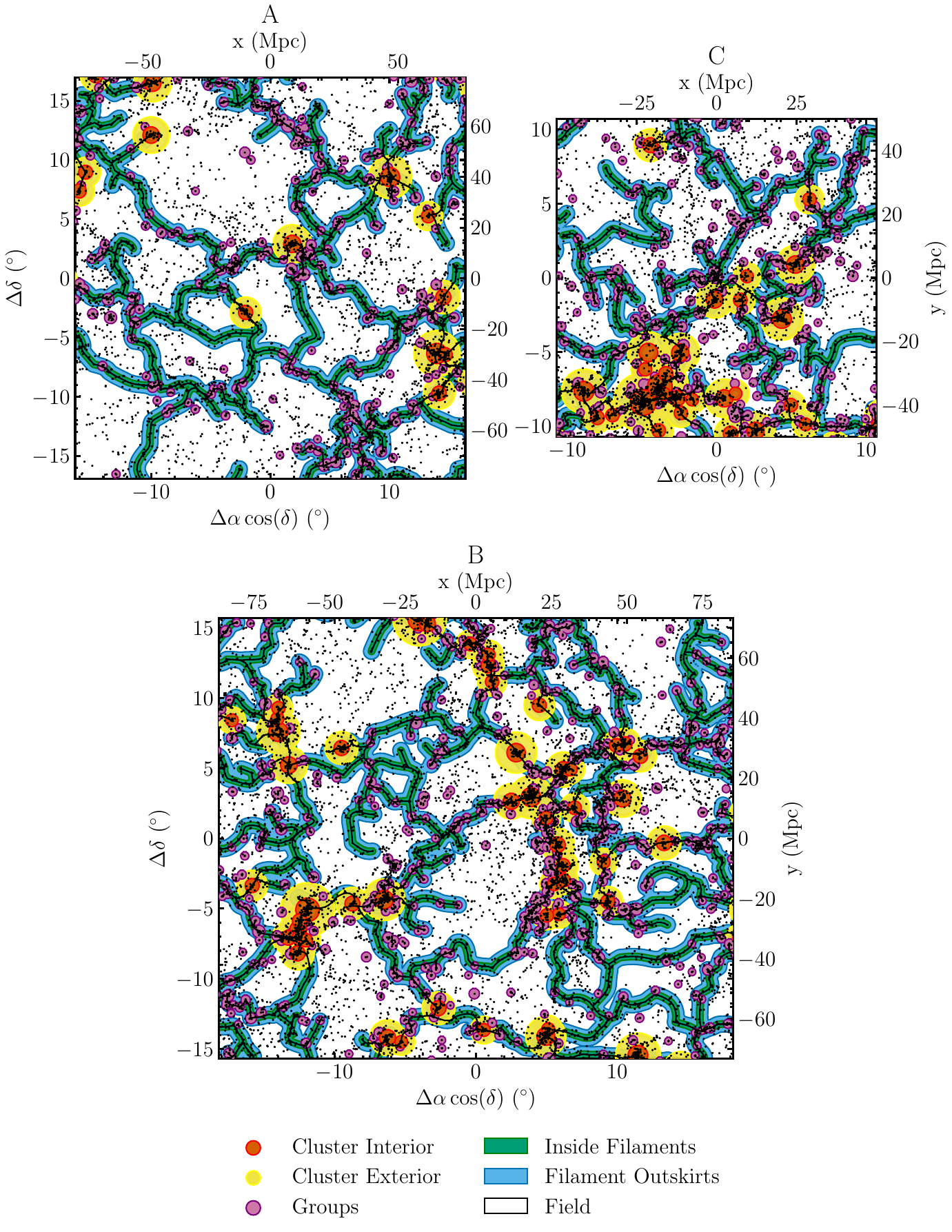}
    \caption{The spatial distribution of galaxies in the mass complete sample considered in this work. Galaxies are shown as black points with \texttt{DisPerSE} filaments as black lines. We identify six distinct environments (see \autoref{tab:Classification_Summary}): Cluster interiors are represented as red circles with radius $R_{180}$ atop yellow circles of radius $2.5R_{180}$, representing cluster 
    exteriors. Groups are displayed as purple circles with radius $R_{180}$. The surface corresponding to filament membership ($D_{\text{fil}} \leq 1\,\text{Mpc}$) is shaded in green. Filament outskirts are similarly shaded in blue, covering the area out to $D_{\text{fil}} = 2.5\,\text{Mpc}$.}
    \label{fig:Spatial_Distribution}
\end{figure*}

\subsection{T-Types}

To extend this investigation to morphologies, we use the catalogue provided by \citet{dominguez_sanchez_improving_2018}, which provides estimates of the galaxy morphologies through the T-Type metric \citep{de_vaucouleurs_revised_1963} for galaxies in the region we consider. These classifications are the result of applying deep learning models trained on 10,000 T-Type morphological classifications in \citet{nair_catalog_2010} to 670,722 SDSS galaxies. In contrast to traditional T-Types, these are not restricted to integers. In this scheme, T-Type$\,<0$ corresponds to early-type morphologies (e.g., E, S0) and T-Types$\,>0$ correspond to late-type morphologies (e.g., Sa and later). Of the 23,441 galaxies in our mass-limited sample, 23,026 (98.2\%) have T-Type estimates.

\section{Methodology} \label{Sec: Methodology}

\subsection{Star-formation suppression} \label{Sec: Star Formation Suppression}

To isolate the effect of the environment on the star-formation activity of the galaxies, we must first account for the variation in star formation rate with stellar mass. To do this, we employ a metric that measures `star-formation suppression', $\Delta$SFMS. This is defined as the vertical logarithmic distance to the Main Sequence of star-forming galaxies in the star-formation rate vs. stellar mass diagram. We identify the Main Sequence, Blue Cloud (BC) and Green Valley (GV) regions by following the work of \citet{trussler_both_2020}
\[\log({SFR}) = 0.7\log({M}_{\text{stellar}}/M_\odot) + b, \tag{1}\label{MS Equation}\]
where $b$ takes values of $-7.52$ and $-8.02$ for the BC/GV and the GV/RS boundary respectively. To define the Main Sequence of star formation, we adopt the value of $b=-7$, as used in \citet{sampaio_blue_2022}. 
A positive value indicates an enhancement in star formation and a negative value indicates a suppression with respect to star-forming galaxies on the Main Sequence. We show the star formation rates as a function of stellar mass in \autoref{fig:SFRvsMass} for the whole galaxy sample, indicating the mass completeness limit as well as the quantity $\Delta$SFMS. This metric of star formation, or similar methods of selecting galaxies based on the distance to the Main Sequence of star formation, has been used successfully in past studies (e.g., \citealt{sampaio_blue_2022}, \citealt{trussler_both_2020}, \citealt{szpila_nature_2024}). In \autoref{fig:SFRvsMass}, we highlight three identifiable regions, The Blue Cloud (BC; $\Delta$SFMS\,$ > -0.52$) containing star-forming galaxies, the Red Sequence (RS; $\Delta$SFMS\,\,$< -1.02$) comprised of quiescent galaxies, as well as the Green Valley (GV; $-1.02 < \Delta$SFMS\,\,$ < -0.52$), containing galaxies transitioning from the BC to the RS.\footnote{The results of this work are unaffected by the exclusion of galaxies with optical signatures of AGN identified through the BPT classifications provided by the MPA--JHU catalogue, outlined in \citet{brinchmann_physical_2004}.}

\subsection{Local densities} \label{sec: Local Densitites}

To investigate effects that may impact galaxy evolution on smaller spatial scales, we employ the projected local galaxy density index $\Sigma^*_3 = {M}_3 / {\pi R}_3^2$, with ${R}_3$ defined as the distance to the $3^{\rm rd}$ nearest galaxy neighbour, following \citet{muldrew_measures_2012}. To probe the local environment, we use mass density, with ${M}_3$ defined as the enclosed stellar mass in a circle of radius ${R}_3$. This definition is comparable to the frequently used number density in system with galaxies of similar mass. It has been shown in the work of \citet{wolf_stages_2009} that number density depends strongly on the selection of the sample, whereas mass density is much more robust to differences in the sample definition.

For galaxies at the edge of the regions, ${R}_3$ may overflow past the region bounds, leading to their densities being underestimated. We identify 787 (3.4\%) galaxies for which this occurs, and they are not included in the subsequent analysis when local densities are required. 

These local galaxy density measurements will be affected by projection effects, where galaxies which are separated a large distance in 3D are included in the density calculation. However, given that the local densities are calculated using the $3^{\rm rd}$ nearest neighbour, ${R}_3$ is relatively small ($\sim1\,$Mpc on average) and so we do not expect that projection effects affect our results.

We emphasise that direct comparisons of local galaxy density indices between different studies is difficult as the resulting densities are dependant on both selection effects and the depth of the redshift window of the sample.

\subsection{Cosmic web extraction} \label{Sec: Cosmic Web Extraction}

Galaxy distributions are expected to trace the cosmic web filaments and we therefore use the galaxies in our sample to map them. Tracing filaments using galaxy distributions is well established with many methods existing to do so. For a list and comparison of some different filament finders, we direct the interested reader to \cite{libeskind_tracing_2018}. A popular method to extract the cosmic web is using \texttt{DisPerSE} (Discrete Persistent Structures Extractor, \citet{sousbie_persistent_2011}). This has been successfully applied in past studies to extract filaments in both 2D and 3D (e.g., \citealt{barsanti_sami_2023,bulichi_how_2023,hasan_how_2023,galarraga-espinosa_evolution_2023,kraljic_galaxy_2018,luber_large-scale_2019,sarron_pre-processing_2019}) and it is the tool we use to identify the filament network in this work.

In our case, we use \texttt{DisPerSE} to identify filaments from the Delaunay tessellation of the given galaxy distribution. The persistent topological features (the critical points maxima, minima and saddle points) of the density field are then identified and filaments are defined as the spatial lines connecting pairs of critical points, i.e., saddle points to the maxima of the density field. A key advantage of using \texttt{DisPerSE} is that it is naturally scale- and parameter-free, whilst allowing for the selection of only the most robust topological features through the persistence ratio. The persistence ratio quantifies the significance of critical point pairs and can be used akin to a signal-to-noise ratio, removing nonphysical features from the filament network. 

The choice of persistence ratio is therefore a compromise between the robustness and the number of filaments. A lower persistence ratio adds more weaker, tendril-like filaments to our sample. As we expect that the environmental effect of filaments is subtle, we want to avoid diluting our filament sample with these less significant filaments, thus compromising our ability to detect the possible effects of filaments. Motivated by this reasoning, we adopt a persistence ratio of 2.5 as this allows us to select the strongest, most significant filaments, as is evident from visual inspection (\autoref{fig:Spatial_Distribution}). Similar values have been used successfully in previous studies (e.g., \citealt{sarron_pre-processing_2019}, \citealt{cornwell_forecasting_2022},  and \citealt{kraljic_galaxy_2018}).

We also include a `5 times' smoothing through the \texttt{-smooth} keyword in the \texttt{skelconv} \texttt{DisPerSE} function. This smooths the network by averaging each point with the coordinates of its neighbour. This is largely an aesthetic choice and does not affect the results of this work. 

With these choices, we are confident to select the dominant filaments responsible for the bulk of galaxy accretion into clusters and therefore relevant for studying pre-processing.

We use the projected 2-dimensional distribution of galaxies to extract filaments. This is because at the redshifts considered in this analysis, $\Delta z = 0.02$ corresponds to $\sim$5600$\,$km$\,$s$^{-1}$ in velocity. The velocity dispersion of a typical galaxy cluster constitutes a sizeable portion of the box depth. Using redshifts as a measure of radial distance could introduce nonphysical spurious filaments tracing the distortions of the Fingers-of-God effect. We refer to the work of \citet{kuchner_cosmic_2021}, which concludes that a 3D cosmic web extraction does not produce reliable filaments in the outskirts of clusters. As a drawback, our choices mean that filaments in this work are largely restricted to those orientated along the plane of the sky. Furthermore, using 2D galaxy positions as tracers for filaments inevitably introduces projection effects. Coherent projection effects result from 3D structures orientated along the line-of-sight. Structures such as walls viewed edge-on would be indistinguishable from filaments; similarly, filaments viewed end-on could resemble clusters. Spurious projection effects are largely random and the result of contamination from foreground and background galaxies. Given that this contamination is expected to be random, and that \texttt{DisPerSE} is relatively robust to noise, the resulting filament network should be largely unaffected. 

We further need to consider unwanted effects introduced close to the edges of our sample area. \texttt{DisPerSE} filaments are subject to edge effects, in the form of spurious filaments tracing the data boundary. A solution would be to extract filaments using the entire main galaxy sample footprint, and then to trim the network to the desired areas of our A, B, C regions. Region C however resides very close to the edge of the footprint and so an alternative approach is required. One solution is that outlined in \citet{cornwell_forecasting_2022}, which involves padding the sample boundary with a random distribution of tracers, with number density equal to that of the sample. We determine this number density using only galaxies further than $2.5R_{180}$ from a cluster centre or $R_{180}$ from a group centre (see section \ref{Sec: Environmental Classifications}). To determine whether this approach is valid, we compare both methods for regions A and B, and find that they produce very similar filament networks.

\begin{table}
    \centering
\caption{Summary of the environmental classification scheme and the ranking of the six distinct environments (see \autoref{fig:Spatial_Distribution}). Galaxies which satisfy multiple classifications are assigned the environment with the highest rank. (see section \ref{sec: Environmental Overlaps})}
    \begin{tabular}{|c|c|c|} \hline 
         Ranking&  Environmental
Classification &          $N$\\ \hline 
         6&  Cluster Interior : $<$ $R_{180, \text{cluster}}$&3105  (13.2\%)\\ \hline 
         5&  Group : $<$ $R_{180, \text{group}}$& 4110  (17.5\%)\\ \hline
         4&  Cluster Exterior : $<$ $2.5R_{180, \text{cluster}}$&1765  (7.5\%)\\ \hline  
         3&  Inside Filaments : $D_{\text{fil}} < 1 \text{Mpc}$ &4905  (20.9\%)\\ \hline 
         2&  Filament Outskirts : $D_{\text{fil}} < 2.5 \text{Mpc}$ &2918  (12.4\%)\\ \hline 
 1& Field : $D_{\text{fil}} > 2.5 \text{Mpc}$ &5251  (22.4\%)\\ \hline
 N/A& Unclassified&1387  (5.9\%)\\\hline
    \end{tabular}
    
    \label{tab:Classification_Summary}
\end{table}

\subsection{Classifications for galaxy environments} \label{Sec: Environmental Classifications}

In this section, we detail the identification of galaxies belonging to any of the six environmental classes highlighted in \autoref{fig:Spatial_Distribution}.

\subsubsection{Inside filaments}

We classify galaxies as members of filaments using the smallest projected distance to a filament $D_{\text{fil}}$. We calculate physical projected distances using the angular diameter distance of the central redshift of the region $z_{\text{centre}}$.\footnote{Throughout this analysis, we determine projected distances using the small triangle approximation.  Over the areas considered in this work, this leads to at most a $\sim$8\% error over $\sim 100\,$Mpc scales relative to the exact treatment. Given its simplicity and computational advantage, we decided to use this approximation since none of our results are affected.}

We model filaments using the skeleton file output of \texttt{DisPerSE}.  This provides a list of segments that make up the filament network. Segments are considered to be linear in angular coordinates ($\Delta \alpha\cos\delta, \Delta \delta$), with $\Delta \alpha = \alpha - \alpha_{\text{centre}}$ and $\Delta \delta = \delta - \delta_{\text{centre}}$. From this, filaments are modelled as a continuous structure, allowing distances to be determined to any point on the filament. 

Past studies have generally found that filaments possess a characteristic radius on the order of $\sim1\,$Mpc \citep{kuchner_mapping_2020, wang_boundary_2024,colberg_inter-cluster_2005,aragon-calvo_multiscale_2010,bond_crawling_2010,cautun_evolution_2014,gonzalez_automated_2010,Castignani_Virgo_2}. We opt to classify galaxies as members of filaments if they reside within 1 Mpc projected distance of a filament. In this scheme, filaments are treated as a homogeneous set with a constant radius. This is an approximation for physical filaments, with their radius and other properties such as density varying as a function of distance to clusters \citep{wang_boundary_2024,pimbblet_intercluster_2004,kim_large-scale_2016, gonzalez_automated_2010,cautun_evolution_2014}. Nevertheless, an approximately constant radius is sufficient for our purposes, and a more sophisticated treatment is not necessary given other uncertainties.

\begin{figure}
	\includegraphics[width=\columnwidth,keepaspectratio]{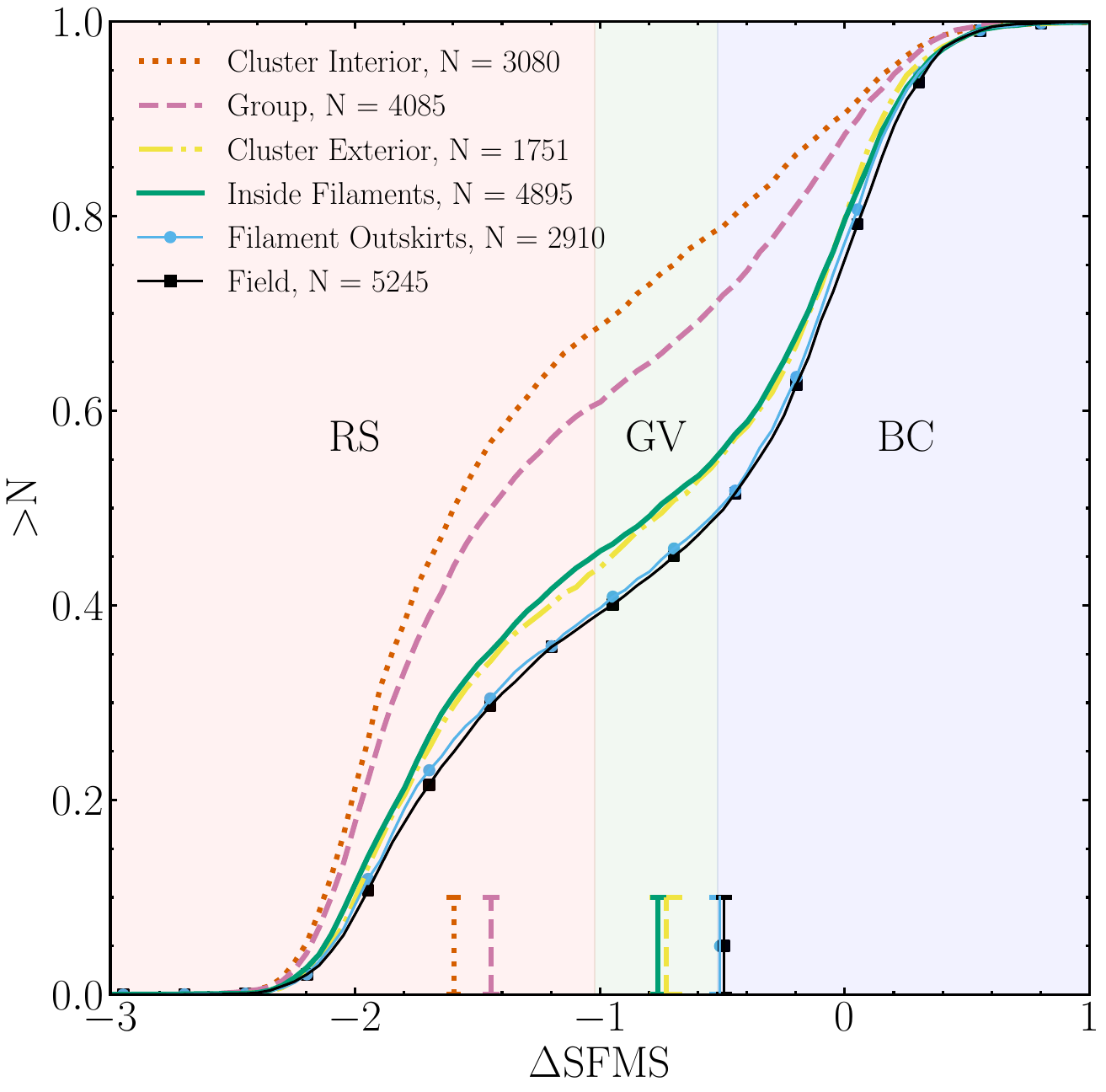}
    \caption{The cumulative distributions of star formation suppression $\Delta$SFMS for the galaxies in each environment. Medians of each distribution are shown as vertical lines along with their respective $1\sigma$ errors. Shaded in red is the region corresponding to the Red Sequence (RS), in green is the Green Valley (GV), and in blue is the Blue Cloud (BC). The numbers of galaxies in each environmental bin are shown in the legend. It is clear from the cumulative distributions that galaxies inside filaments tend to be suppressed in star formation relative to those in the field, whilst galaxies in groups and cluster interiors tend to be suppressed in star formation relative to those in filaments. This suggests that filaments may act as an intermediary environment.}
    \label{fig:SFMS_Cumulative}
\end{figure}

\subsubsection{Filament outskirts}

Filaments are high-vorticity structures, with both gas and dark matter profiles known to be well described by a self-gravitating isothermal profile \citep{lu_structure_2023,ramsoy_rivers_2021}. There is however evidence to suggest that filaments are also highly complex. One such example is the one presented by \citet{lu_structure_2023}, which investigates the radial profile of three simulated Mpc-scale filaments at $z\sim 4$. They found that the radial profile of filaments can be described with three zones: an inner zone with cold dense gas, an intermediary zone dominated by vortices due to inflowing and post-shock gas, and an outer zone where outwards thermal pressure decelerates inflowing gas. While the structure of filaments at $z\sim0$ is currently not well understood, \cite{song_beyond_2021} provides evidence that at $z \sim 2$, the distribution of halos around filaments is bimodal, with some galaxies very close to filament and others further away ($\sim 1$ Mpc). The authors find that the efficiency of galaxy mass assembly is specific to the distance to a filament spine, with galaxies at the edge of filaments potentially subject to unique quenching mechanisms. This suggests that a simple description of filament membership or not is insufficient and motivates the inclusion of a secondary filament environment, the filament outskirts. This environment further aids our investigation by increasing the contrast between the filament and field populations, allowing for a `buffer zone' separating `true' filament galaxies from field galaxies that are most likely not to have interacted with the higher density regions of the cosmic web. We classify galaxies as members of filament outskirts if they satisfy $1 \text{Mpc} < {D}_{\text{fil}} < 2.5 \text{Mpc}$. 

\subsubsection{Field}

Galaxies in the field form our control sample. These are galaxies which exist in low-density regions and are unlikely to be influenced by collapsed elements of the cosmic web e.g. walls, filaments and nodes. We classify galaxies as members of the field if they satisfy ${D}_{\text{fil}} > 2.5 \text{Mpc}$ and are not members of higher-density environments such as groups, clusters, and their outskirts (see below). 

\begin{figure}
	\includegraphics[width=0.99\columnwidth,keepaspectratio]{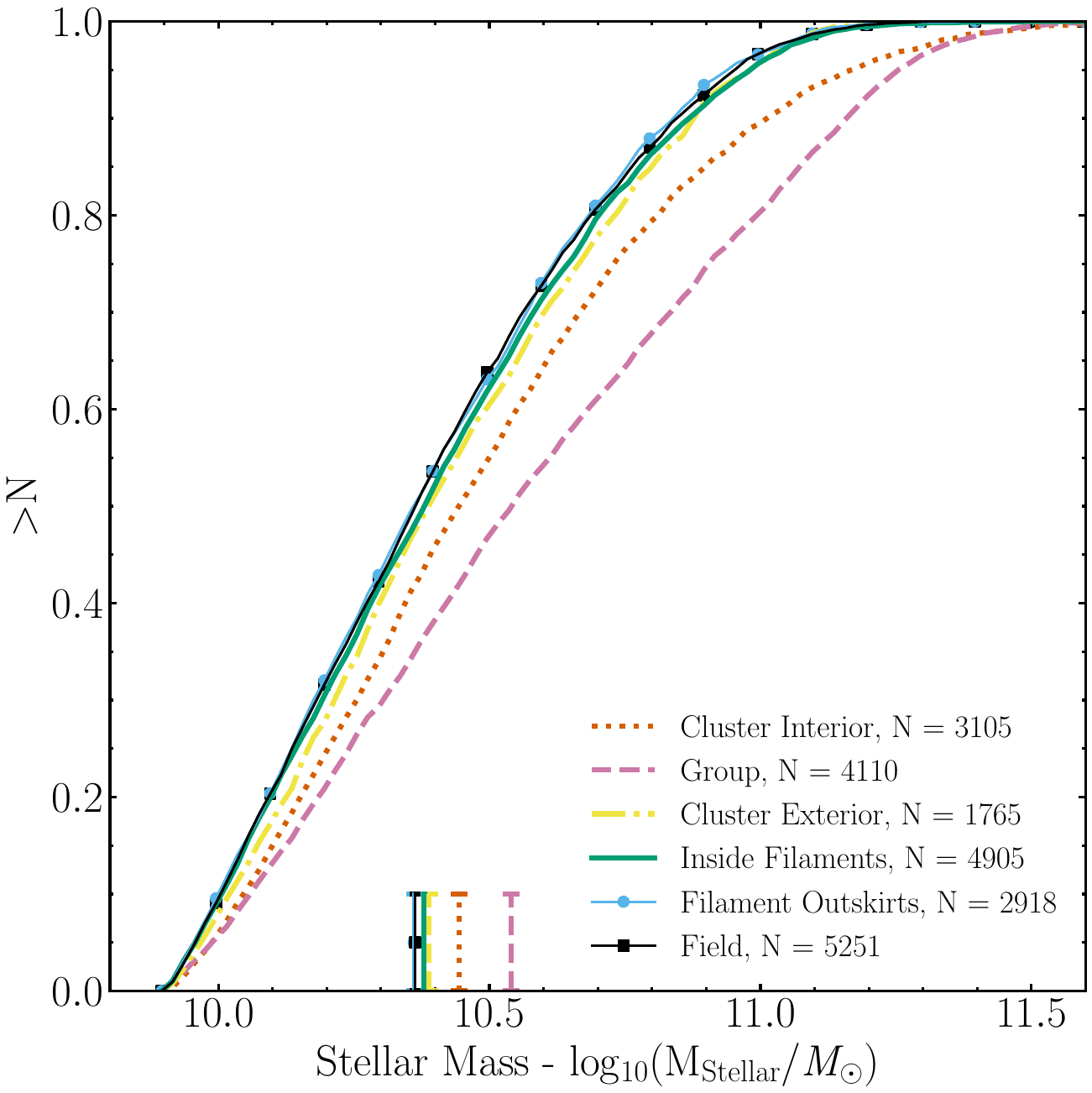}
    \caption{The cumulative distributions of galaxy stellar mass for the galaxy populations in each environmental bin. Medians for each distribution are shown by vertical lines along with their respective $1\sigma$ errors. The numbers of galaxies in each environment are shown in the legend. It is clear that `denser' environments tend to posses galaxies with higher stellar masses. Curiously, we find that galaxies in groups tend to be of higher masses than those in cluster interiors, a trend which is discussed in section \ref{Results: Stellar Mass}.}
    \label{fig:Mass_Cumulative}
\end{figure}

\subsubsection{Cluster and groups} \label{Methods: Clusters and Groups}

Galaxies which have been affected by the cluster and group environments must be removed from our sample of filament members if we are to isolate the effect of the filament environment itself. These galaxies may have already undergone some transformation, which could erroneously be interpreted as the effect of filaments if they are not removed from the filament galaxy sample. This exclusion is commonplace in the literature; one example is presented in the work of \citet{laigle_cosmos2015_2018}, in which galaxies closer than some distance to a cluster/group are excluded from the sample. We adopt a similar approach. We use the $R_{180}$ estimates provided for groups and clusters in the YGC, where $R_{180}$ is the radius of a sphere whose mean density is 180 times the critical density. We determine $R_{180}$ using equation 5 in \citet{yang_galaxy_2007}.\footnote{$R_{180}$ is somewhat larger than the commonly-used $R_{200}$, and therefore provides a more conservative exclusion zone around clusters and groups. Moreover, since the YGC provides $R_{180}$ directly, we prefer to use this value rather $R_{200}$ to avoid model-depending scaling. }

\begin{figure*}
	\includegraphics[width=1.8\columnwidth,keepaspectratio]{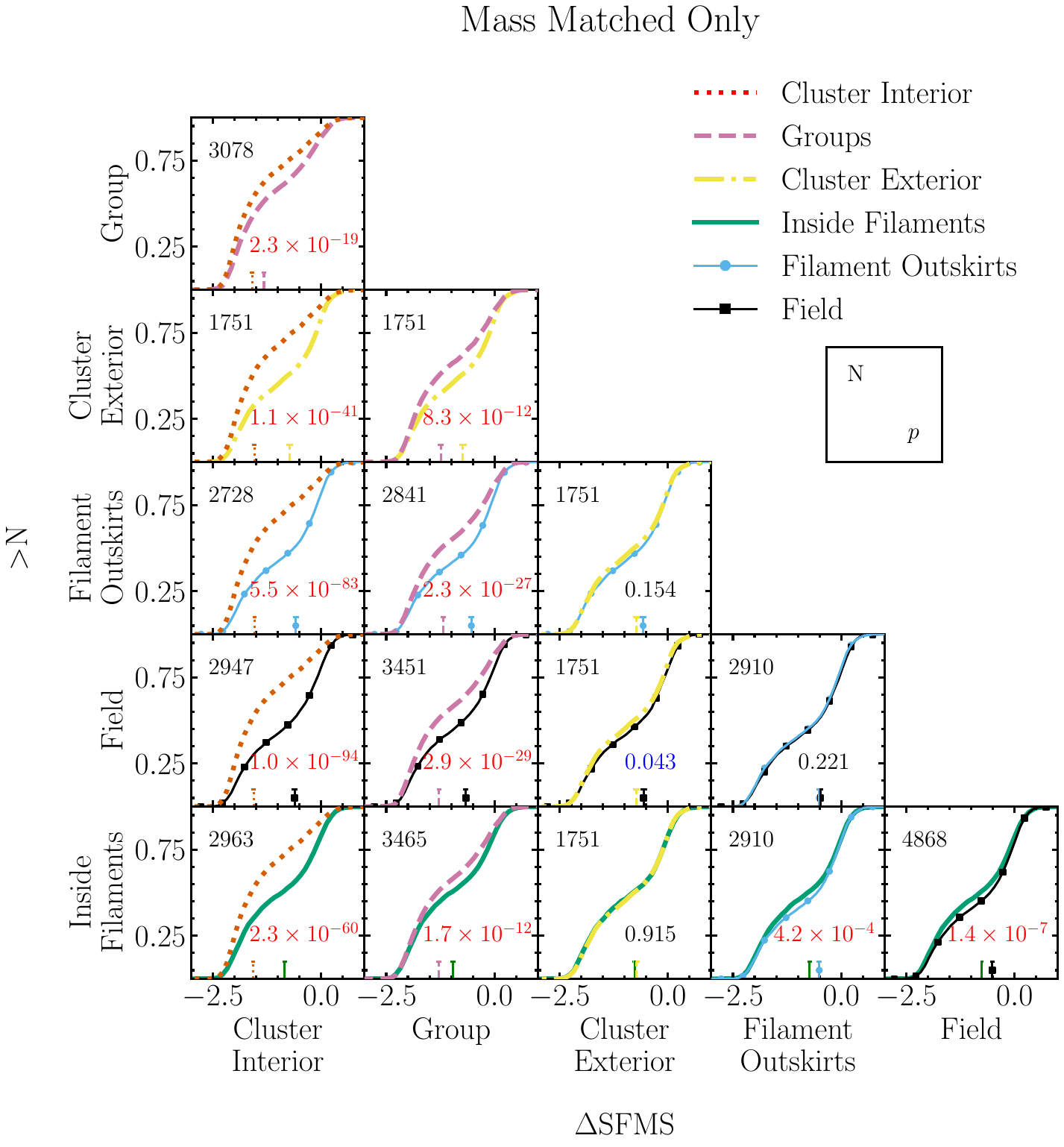}
    \caption{Pairwise comparisons of the $\Delta$SFMS cumulative distributions for each environmental pair using the mass-matched samples only. Medians of each distribution are shown as vertical lines with their respective $1\sigma$ errors. In each panel, the number of galaxies in each population is shown in the top left. Using Kolmogorov-Smirnov statistics, the probability that both distributions are identical is shown in the lower right. Significant p-values ($p < 0.05$) are coloured in blue and highly significant p-values ($p < 0.01$) are coloured in red. The comparisons between the distributions of galaxies inside filaments and those within cluster interiors and the field, show highly significant differences ($p < 0.01$). Further showing that when matching in stellar mass, filaments appear as an intermediary environment between the clusters and the field.}
    \label{fig:MassOnlyPairwiseSFMS}
\end{figure*}

We classify galaxies as members of clusters/groups using the projected distances to a cluster/group centre, in combination with the velocity dispersion of the cluster/group. We determine the line-of-sight velocity dispersion using equation~6 in \citet{yang_galaxy_2007}. Galaxies within projected $R_{180}$ of a cluster/group centre, and with redshifts such that $|z - z_{\text{cluster/croup}}| < 3z_{\sigma}$, are classed as members of the group/cluster interiors. These classifications of group and cluster interior galaxies are similar to what is often refereed as `cluster members' and `group members' in previous works. We note that the results presented in this paper do not change if we assign memberships entirely in 2D, defining cluster/group members as those galaxies inside a cylinder of radius $R_{180}$ and height $\Delta z = 0.02$.

For galaxies within this projected distance, but with redshift exceeding $\pm 3z_{\sigma}$, it is not clear what environment these belong to. As such, of the 23,441 galaxies in our sample, 1,387  (5.9\%) galaxies are not given any classification and are not considered in the following analysis comparing galaxies in different environments.

This treatment is not sufficient to remove the influence of clusters due to the existence of `backsplash' galaxies. These are defined in the literature as galaxies which have been within $R_{200}$ of a cluster at some point in the past but are now located further away \citep{gill_evolution_2005, bahe_why_2013}. Using observational data it is not possible to definitively identify individual backsplash galaxies; only the probability that a galaxy is backsplash can be determined. \citet{kuchner_inventory_2021} found that, for clusters, at $R_{200}$ $30$--$60\%$ of galaxies within filaments are likely to be backsplash, with the exact number dependant on the cluster dynamical state. This probability vanishes at $\sim2.5R_{200}$. To account for these backsplash galaxies, we classify galaxies as members of the `cluster outskirts' if they are within $R_{180}$ and $2.5R_{180}$ of a cluster centre. Note that we still use $R_{180}$ in this work as a slightly larger and therefore more conservative estimate of the clusters' sphere of influence than $R_{200}$.

Backsplash galaxies in groups are expected to be fewer than those of clusters, resulting from their reduced richness and gravitational potential. As a result, we do not attempt to correct for backsplash galaxies in groups. Furthermore, since groups are overwhelmingly located in filaments, removing galaxies within $2.5R_{180}$ of groups unnecessarily decreases our filament galaxy sample.

\subsection{Environmental overlaps} \label{sec: Environmental Overlaps}

Many galaxies in our sample satisfy more than one of the classification criteria. We therefore assign a hierarchy of classifications, to prevent overlaps. When assigning a galaxy to an environmental class, if it meets the criteria for more than one class, it is allocated to the class with the higher rank (`denser' environment). This ranking is presented in \autoref{tab:Classification_Summary}. This is to avoid mixing high and medium density environments in an attempt to separate their associated environmental effects (for a detailed discussion of the environmental effects in high density environments, particularly the difference between those in clusters and groups, we direct the interested reader to the recent review of \citealt{alberts_clusters_2022}). It is shown in Table~4 of \citet{aragon-calvo_multiscale_2010} that different components of the cosmic web possess a characteristic density, with clusters denser than filaments, which are in turn more dense than voids. We expect that clusters should be of a higher density than groups, due to their increased richness and halo mass. Using the reasoning above, the cluster outskirts must be ranked higher than filaments to account for backsplash galaxies. Galaxies often fall onto clusters as part of groups, therefore we opt to rank groups higher than the cluster outskirts to retain these infalling groups as part of the group class. 

\section{Results and discussion} \label{sec: Results}

In this section we first compare the cumulative distributions of galaxy properties within each environment. We then repeat the analysis but accounting for differences in stellar mass distributions across environments by matching the galaxy samples in mass. Finally, we account for differences in the local galaxy density distributions and present the results obtained with samples matched in both stellar mass and local galaxy density. We present the results for the star-formation suppression index $\Delta$SFMS in section~\ref{Sec: Results - SFMS} first and for the galaxy morphologies in section~\ref{subsection: Results - T-Types}

\subsection{Star-formation suppression} \label{Sec: Results - SFMS}

\subsubsection{Full galaxy samples} \label{Results: Full galaxy sample}

Galaxies in groups and clusters have suppressed star formation compared to those in filament galaxies, which are in turn more suppressed that field galaxies (\autoref{fig:SFMS_Cumulative}). This is evident from their cumulative distributions of $\Delta$SFMS. Whilst this immediately suggests the presence of environmental effects and, in particular, pre-processing in filaments, there is strong evidence that stellar mass is a key factor driving the properties of galaxies (e.g., \citealt{alpaslan_galaxy_2015,oesch_buildup_2010}). As such, to ensure that the effects observed are truly a consequence of the environment and not due to differences in mass, we need to compare the stellar mass distributions between the galaxy populations across environments. 

\subsubsection{Stellar mass distribution} \label{Results: Stellar Mass}

To ascertain the effects of stellar mass on the observed trends in $\Delta$SFMS, we investigate how the stellar mass functions are correlated with the environment. For the galaxies in each environmental class, we present the cumulative mass distributions in \autoref{fig:Mass_Cumulative}. It is clear that the distributions are not equal. Cluster interiors and groups tend to be significantly more massive than those in filaments. We also find hints of a difference between the stellar mass distributions of galaxies in filaments and those in the field, this difference is small and statistically insignificant ($p = 0.149$).

\begin{figure*}
	\includegraphics[width=1.8\columnwidth,keepaspectratio]{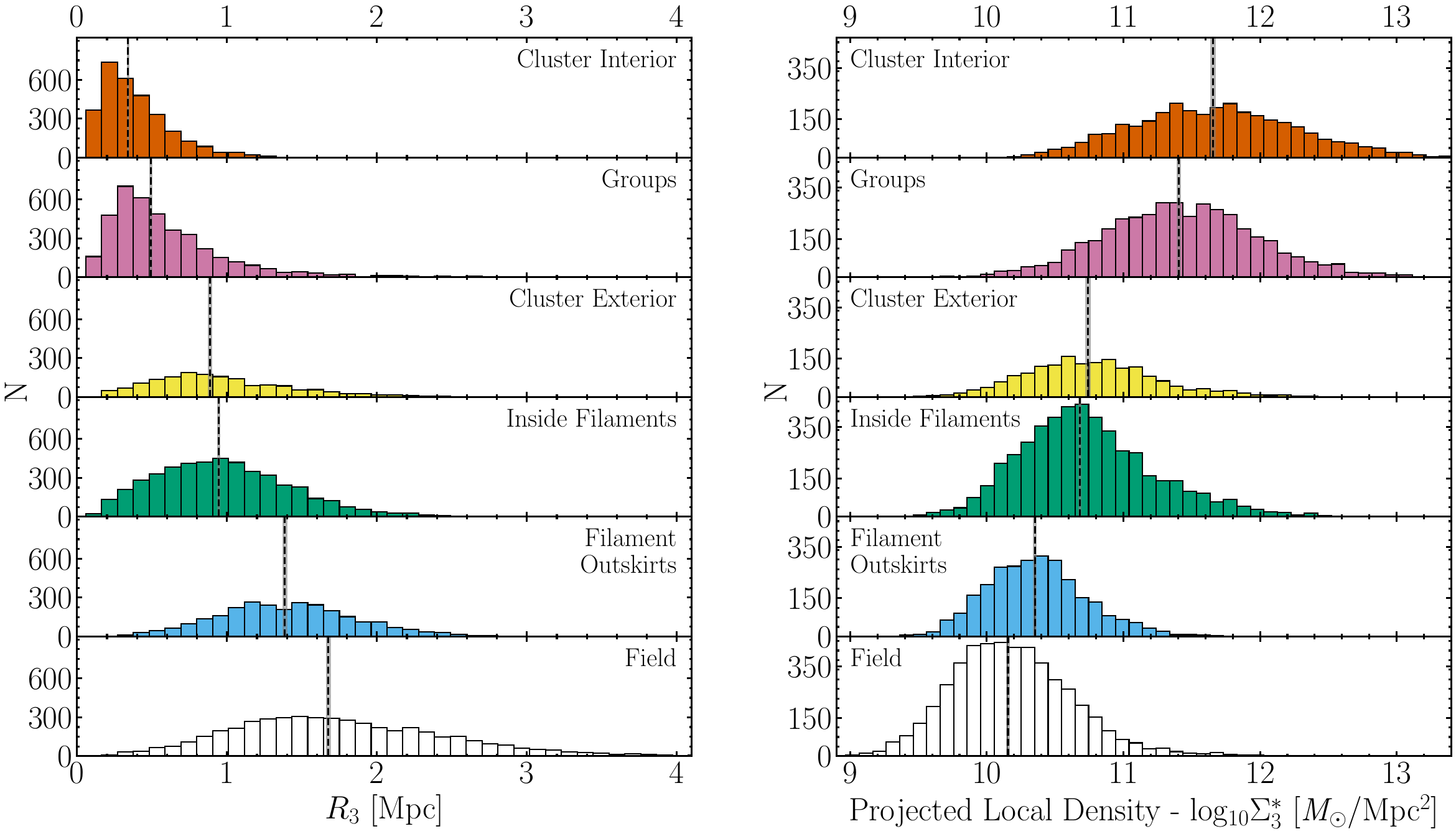}
    \caption{(Left column) Distributions of the distance to the third nearest neighbour $R_3$. (Right column) Distributions of the projected local galaxy density index $\Sigma^*_3$. Each panel corresponds to a separate environmental class, coloured consistent with the other figures in this work. Medians of each of the distributions are shown by vertical dashed lines as well as the $1\sigma$ errors shown by the shaded gray area. The environments considered in this work posses different, yet overlapping $R_3$ and local density distributions, highlighting both the multiscale nature of the cosmic web and the need to consider both the local and large scale environment. The median value of $R_3$ in filaments is $\sim 1$\,Mpc, half the width of the filaments considered in this work. This shows that $R_3$ is an ideal choice to probe the internal environment of filaments as it tends to probe the the density inside filaments themselves. This is also the case for other environments such as groups and cluster interiors -- $R_3$ measures densities \textit{inside} each specific environment.
    }
    \label{fig:Density_Distributions}
\end{figure*}

Galaxies near filaments tending to be more massive than those further away is a well-established result in the literature. Mass gradients have been reported in many prior studies, examples include \citet{sarron_pre-processing_2019,malavasi_vimos_2017,kraljic_galaxy_2018,bulichi_how_2023, ricciardelli_morphological_2017,hoosain_effect_2024,alpaslan_galaxy_2015,song_beyond_2021}. This trend can be explained by an enhancement of merger rates inside filaments \citep{malavasi_vimos_2017}. Alternatively, this could be the results of the biasing of the mass function around the large-scale structure (Kaiser Bias; \citealt{kaiser_spatial_1984}), who propose that the enhanced density field facilitates an earlier collapse of proto-halos, leading to an excess of massive galaxies in denser environments.

Surprisingly, we find that galaxies in groups tend to be of higher masses than galaxies in clusters. The effect is small, with the median stellar mass $0.1\,$dex ($26\%$) higher in groups than in cluster interiors (a $6.75\sigma$ difference). This result contradicts that of \citet{alpaslan_galaxy_2015}, who investigated how the galaxy stellar mass functions vary as a function of cosmic web environment using GAMA data. Their galaxy stellar mass functions for high- and mid-mass groups correspond to those of our clusters and groups respectively. Contrary to our findings, \citeauthor{alpaslan_galaxy_2015} find that their high-mass groups (corresponding to our clusters) generally contain a larger proportion of high-mass galaxies than their mid-mass groups (what we call `group'). We do not have a clear explanation for this contradiction but, given the relatively small size of the effect and the differences in sample selection and mass determination, it is perhaps not too surprising that our results do not match exactly. 

It is nevertheless clear that differences in the stellar mass distributions may be responsible for the trends observed in \autoref{fig:SFMS_Cumulative}. To account for this, we will construct mass-matched samples (i.e. samples with identical stellar mass distributions).

\subsubsection{$\Delta$SFMS -- mass matched sample}

We construct mass-matched samples through pairwise comparisons of each environment. We construct these mass-matched samples as follows. For each environment pair (e.g., inside filaments and field), we take each galaxy within the class with the smallest number of galaxies, and find its pair in the other with the closest mass. If this mass is within $0.1\,$dex, then both galaxies are added to their respective mass-matched samples. In this procedure, we do not allow for replacement.

We present the results of this pairwise comparison in \autoref{fig:MassOnlyPairwiseSFMS}. We find that the trends shown in \autoref{fig:SFMS_Cumulative} are also evident in the mass-matched pairwise comparisons. We find that when mass-matched, galaxies inside filaments tend to be suppressed in star formation relative to the field population and further suppressed relative to galaxies in groups and within the interiors of clusters. This supports the argument that filaments may be an intermediary environment, between the field and clusters. 

Interestingly, we find that the $\Delta$SFMS distributions for filaments and cluster outskirts are similar (both for the mass-matched samples and the un-matched ones). In hindsight, this should not be surprising -- based on simulations, \citet{kuchner_inventory_2021} show that the outskirts of clusters are highly heterogeneous environments, with up to 45\% of galaxies in this region closer than $1h^{-1}\,$Mpc to a filament spine, with 28--58\% (dependant on cluster dynamic state) unaffected by the cluster. As such, the population of the cluster outskirts is expected to consist mainly of filament, and field galaxies and therefore it is expected that galaxies in the cluster outskirts show little difference from the filament and field populations.

The increased star-formation suppression within filaments, in comparison with the field, agrees with the results of past studies such as \citet{kraljic_galaxy_2018, laigle_cosmos2015_2018, martinez_galaxies_2016}, as well as simulation work (e.g.,  \citealt{bulichi_how_2023}). The increased passive fraction and suppressed star formation of filament galaxies relative to the field can be interpreted as galaxies undergoing some degree of pre-processing inside filaments before they infall into clusters.

\begin{figure*}
	\includegraphics[width=1.8\columnwidth,keepaspectratio]{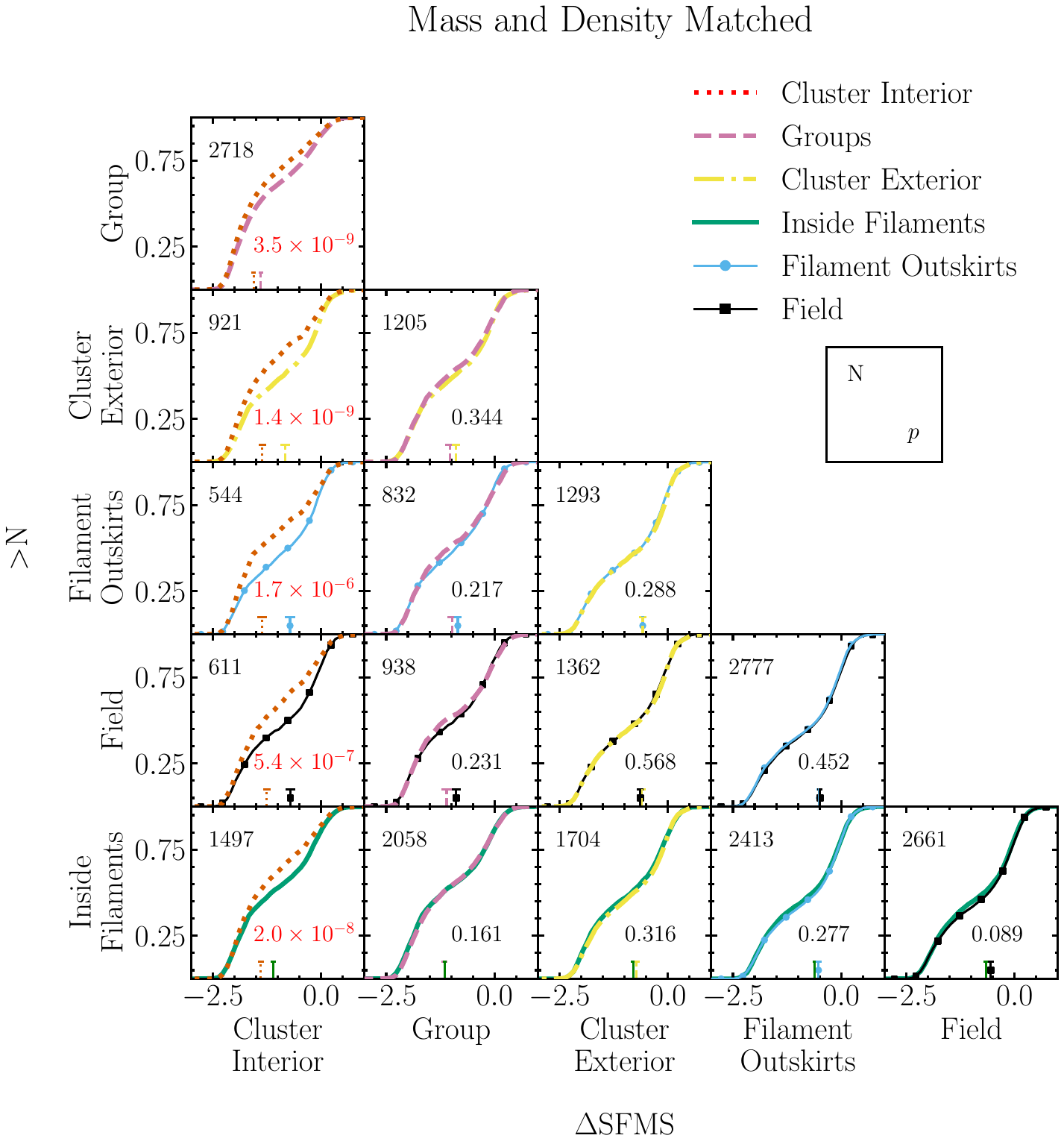}
    \caption{Pairwise comparisons of the $\Delta$SFMS distributions for each environmental pair using the mass and local galaxy density matched samples. Medians of each distribution are shown as vertical lines with their respective $1\sigma$ errors. In each panel, the number of galaxies in each population is shown in the top left. Using Kolmogorov-Smirnov statistics, the probability that both distributions are identical is shown in the lower right. Significant p-values ($p < 0.05$) are coloured in blue and highly significant p-values ($p < 0.01$) are coloured in red. Whilst the differences between the filament population observed in the stellar mass matched samples (\autoref{fig:MassOnlyPairwiseSFMS}) persist between filaments and cluster interiors, the difference between the filament population and the field vanishes. This suggests that the environmental effect on $\Delta$SFMS of galaxies within filaments can entirely be characterised by a local galaxy density index $\Sigma^*_3$.}
    \label{fig:PairwiseSFMS}
\end{figure*}

\medskip

The analysis of the effect of filaments on star formation with mass-matched samples is commonplace in the literature. However, while this indicates that the filamentary environment does indeed play a role in shaping galaxy properties, it is not clear if these differences in $\Delta$SFMS are driven by the physical processes associated with the small-scale or the large-scale environment. By parameterising the local environment of a galaxy through the local galaxy density index $\Sigma^*_3$ (\autoref{fig:Density_Distributions}), we attempt to gain insight into this by constructing samples matched in both stellar mass and local galaxy density.

\subsubsection{$\Delta$SFMS -- mass and local-density matched samples} \label{subsubsection: SFMS}

The distributions of $\Sigma^*_3$ for the different environments are shown in \autoref{fig:Density_Distributions}. Although the range of densities present in each environment is quite diverse, there is enough overlap to build density-matched galaxy samples for all pairs of environments. 

Following on what we did before, we now carry out pairwise comparisons between each environmental bin with samples matched both in stellar mass and local galaxy density (i.e. Identical stellar mass and local galaxy density distributions). To do this, for each pair of environments, we take the sample with the fewest galaxies and, for each galaxy, we collect all the galaxies in the other environment with stellar mass within $0.1\,$dex. Of these we select the closest in local galaxy density. If this pair has a local galaxy density within $0.1\,$dex, then both are added to the matched sample. We do not allow for replacement.

Many of the differences which were present in the mass-matched comparison vanish when also matching in local galaxy density (\autoref{fig:PairwiseSFMS}). We find that differences remain only for comparisons with cluster interiors. Interestingly, we find that the $\Delta$SFMS distribution for the filament populations appears statistically indistinguishable from that of the field population when matching in stellar mass and local galaxy density.

\begin{figure}
	\includegraphics[width=\columnwidth,keepaspectratio]{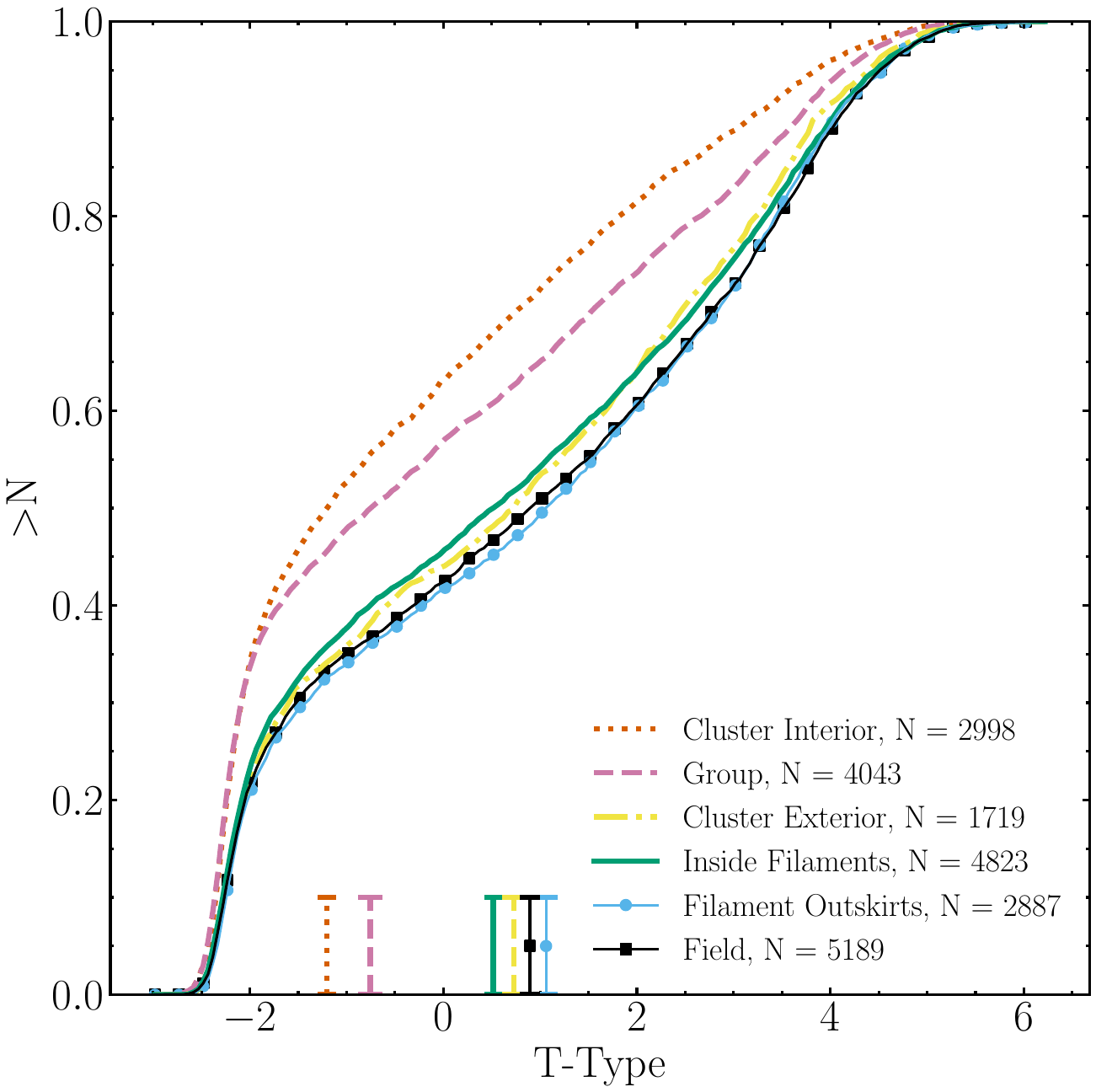}
    \caption{Cumulative distributions of T-Types for the galaxy populations in each environmental bin. Medians of each distribution are shown as vertical lines along with their respective $1\sigma$ errors. As with $\Delta$SFMS, we find that galaxies inside filaments tend to favour more early-type morphologies than those in the field, whilst galaxies in groups and cluster interiors tend to posses more early-type morphologies than those in filaments. This suggests that filaments act as intermediary environments for morphological transformations.}
    \label{fig:TTYPE_Cumulative}
\end{figure}

Given the differences (or lack thereof) found in the pairwise comparisons of the $\Delta$SFMS distributions when simultaneously matching in stellar mass and local galaxy density, we conclude that the effect of filaments on $\Delta$SFMS can be entirely parameterised by a local galaxy density index $\Sigma^*_3$. In other words, galaxies in filaments are subject to environmental processes that correlate well with local galaxy density.

Whilst it is not possible to conclude what these processes are with the available information, we can speculate what these could be. One natural choice is an enhanced rate of galaxy-galaxy interactions; this environmental effect will be elevated inside filaments due to the increased number density of galaxies relative to the field. Local galaxy density acts as a proxy for such effects and could explain the differences seen between the mass-matched samples, but not seen in the mass- and local galaxy density-matched ones. This scenario would be consistent with the results and previous works discussed in section \ref{Results: Stellar Mass} which find that filament galaxies tend to have a greater stellar mass compared to those in the field, perhaps hinting at an enhanced merger rate inside filaments.

Additionally, whilst the difference observed in the mass-matched samples may be due to environmental processes governed by local galaxy density, it could also be that the relevant physics simply \textit{correlates} with it. One such possibility could be that the differences are a consequence of elevated galaxy halo masses at fixed stellar masses. \citet{oyarzun_sdss-iv_2023} find that galaxy halo mass is an important parameter in describing the properties of passive satellite galaxies. Similar results are found by \citet{scholz-diaz_dark_2022} for central galaxies and by \cite{zhou_separate_2024} for central disk galaxies. Simulations such as those of \citet{wang_environmental_2023} also point towards the importance of halo mass in driving galaxy evolution, even at fixed stellar mass. There is evidence that galaxy halo mass is correlated with stellar mass \citep{wu_how_2024} and local density \citep{haas_disentangling_2012,muldrew_measures_2012}. While the extent of the correlation between galaxy halo mass and local density is disputed, especially for low galaxy halo masses ($\log(M/\text{M}_{\odot}) < 13$), it is possible that the differences seen in \autoref{fig:MassOnlyPairwiseSFMS} could be due to the difference in galaxy halo mass distributions in the stellar mass-matched samples across environments. It is too early to pinpoint the exact effect these differences in halo masses would have on specific galaxy properties such as $\Delta$SFMS, but there is a growing body of empirical and theoretical evidence suggesting that it may be important. 

Another possibility is that these differences are a consequence of ``archaeological downsizing'' \citep{thomas_epochs_2005}.  Galaxies residing in overdense regions likely formed before those in underdense regions, with the enhanced density field facilitating an earlier proto-halo collapse. Given that these galaxies will have begun assembling their stellar populations earlier, and mass quenching is strongly mass dependant with higher mass galaxies quenching first (e.g. \citealt{darvish_effects_2016,popesso_effect_2011,sobral_dependence_2011}), it is expected that galaxies in dense environments such as clusters, groups and filaments, should have an excess of older galaxies relative to those in the field. Therefore, it is also possible that the observed suppression in star formation of filament galaxies is at least in part a result of older stellar ages. Given that we do not have age estimates for the galaxies in this work, we cannot investigate the extent to which differing ages could contribute to these observed differences.

While most of the differences in the $\Delta$SFMS distributions across environments disappear when we match both in stellar mass and local galaxy density, some significant differences persist -- albeit at a small level -- when comparing other environments with cluster interiors. This suggests that even if part of the environmental effect of cluster interiors can be characterised by local galaxy density, there are additional physical processes not directly correlated with it. This means that, in many ways, these are unique environments where galaxies are subject to physical processes beyond those linked with local galaxy density. One candidate is ram-pressure stripping by the ICM. There is strong evidence that galaxies infalling into galaxies clusters show signatures of ram pressure stripping (e.g. \citealt{vulcani_relevance_2022,poggianti_gasp_2017,vulcani_gasp_2020}). However, whether or not signatures of ram pressure stripping in filaments are to be expected, is not entirely known. In the work of \citet{thompson_gas_2023} a set of void simulations are used to investigate the ability of haloes to accrete gas in voids, finding that even in low density void walls, ram pressure stripping can occur, impairing the accretion of gas. Furthermore, \cite{song_beyond_2021} find using the HORIZON-AGN simulations that the high vorticity regions of filament edges could reduce the efficiency of gas transfer within galaxies due to the coherent and large angular momentum of the outer halo as fed by these vorticity rich filaments. The work of \cite{kotecha_cosmic_2022} using The Three Hundred Project simulations suggests that filaments may even shield galaxies from the ICM and limit ram pressure stripping, with cluster galaxies near filaments tending to be more star forming than those further away. As discussed in \citet{darvish_cosmic_2014}, ram-pressure stripping is not expected to be an effective mechanism to suppress star formation within filaments due to the reduced density of the inter-galactic medium relative to the ICM, together with the smaller velocities of filament galaxies compared with those inside clusters. Together with the trends shown in \autoref{fig:PairwiseSFMS}, this suggests that clusters may be extreme and unique environments with additional environmental effects not experienced by galaxies in filaments alone. 

We again emphasise that given the available information, we can only conclude that $\Sigma^*_3$ encodes the differences observed in the mass-matched samples of filaments and field. The above discussion  concerning the possible mechanisms and physical processes is reasonable and plausible, but speculative. We do not yet have the required information to make firmer conclusions. As we discuss in \autoref{sec: Conclusions}, exploring the star-forming histories of galaxies living in different environments will help us to make progress.

We cannot end this discussion concerning galaxy densities without pointing out that the scale at which one computes galaxy density probes environment and the related physics on different spatial scales. Small scales probe the local environment and the most recent and stochastic processes, whereas larger scales take into account the integrated -- and thus smoother -- environmental history of the galaxies. Moreover, since using $n=3$ when computing $\Sigma^*_n$ is somewhat arbitrary, we checked that our results persist for $n = 5$. Finally, even though we argued in section~\ref{sec: Local Densitites} that using $\Sigma^*_n = {M}_n / {\pi R}_n^2$ makes more physical sense that using $\Sigma_n = n / {\pi R}_n^2$ (ignoring the mass of the companion galaxies), we checked that using $\Sigma_3$ instead of $\Sigma^*_3$ leaves our results largely unchanged also. 

\subsection{T-Types} \label{subsection: Results - T-Types}

We now extend our analysis to galaxy morphologies using T-Types. We present the T-Type cumulative distributions for the whole sample in \autoref{fig:TTYPE_Cumulative}. We observe similar trends as in $\Delta$SFMS (cf. \autoref{fig:SFMS_Cumulative}), where the cluster interior and group environments are significantly biased towards earlier-type morphologies relative to the other environments. We also find a difference between the filament population and the field, in which filament galaxies preferentially exhibit more early-type morphologies than field ones. This suggests that filaments do contribute, at least to some extent, to the morphological transformation of galaxies. However, as before, we must match our samples in mass before any inferences can be made about the specific effect on the environment.

\begin{figure*}
	\includegraphics[width=1.8\columnwidth,keepaspectratio]{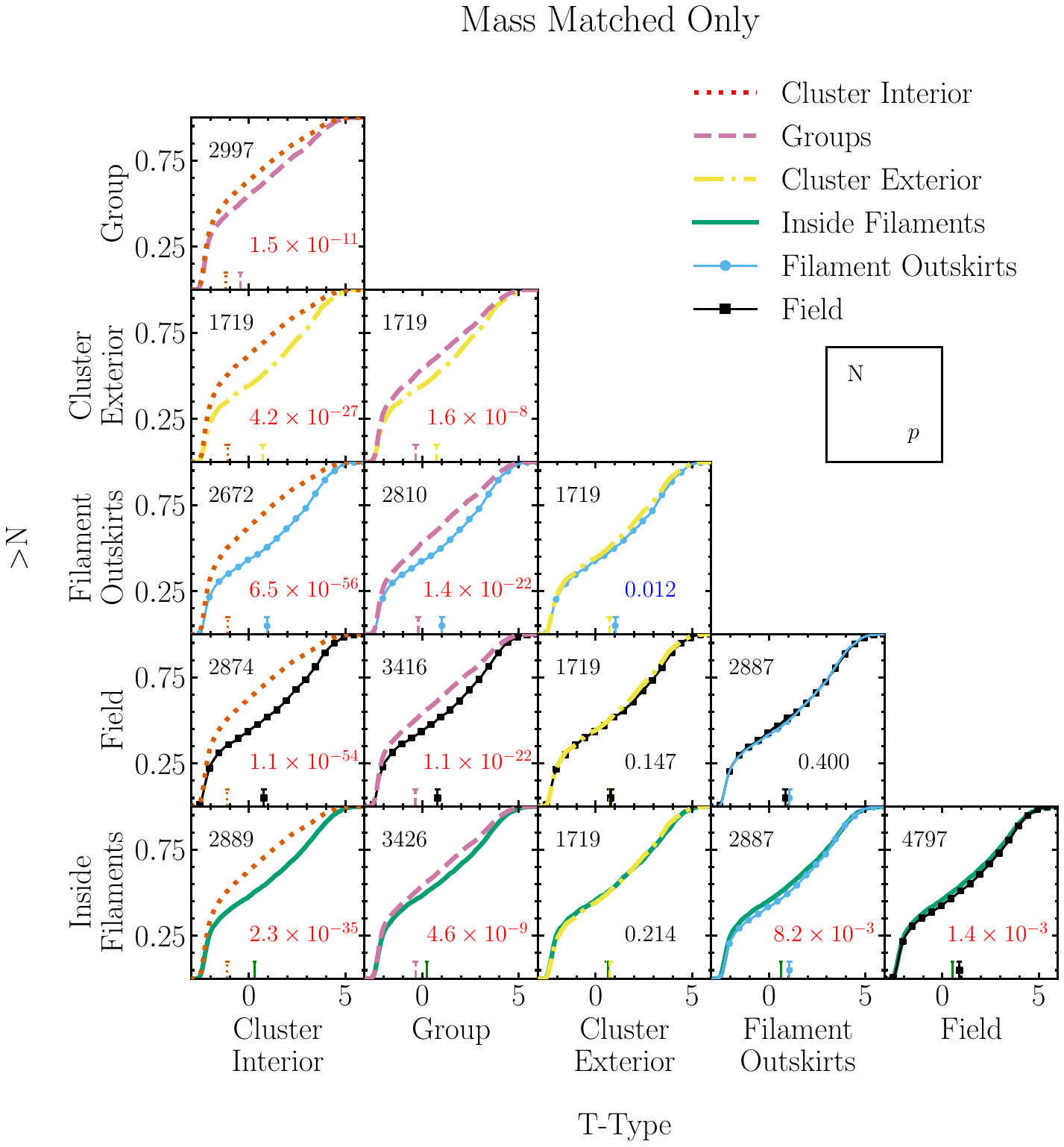}
    \caption{Pairwise comparisons of the T-Type distributions for each environmental bin pair, using the mass-matched samples only. Medians for each distribution are shown as vertical lines along with their respective $1\sigma$ errors. In each panel, the number of galaxies in each population is shown in the top left. Using Kolmogorov-Smirnov statistics, the probability that both distributions are identical is shown in the lower right. Significant p-values ($p < 0.05$) are coloured in blue and highly significant p-values ($p < 0.01$) are coloured in red. There are significant differences between the T-Type distributions between the mass matched samples of galaxies within filaments compared to those in cluster interiors and the field. This suggests that when matching in mass only, filaments act as an intermediate environment between clusters and the field.}
    \label{fig:MassOnlyPairwiseTType}
\end{figure*}

\subsubsection{T-Type -- mass matched samples}

We present the pairwise comparisons of the mass-matched samples in \autoref{fig:MassOnlyPairwiseTType}. We find that there appears to a be small yet statistically significant difference between the T-Type distributions of the filament and field populations ($p = 1.4\times10^{-3}$). We find that galaxies in filaments tend to be slightly more early-type in morphology than those in the field, suggesting that galaxies in filaments may be subject to pre-processing not only affecting their star-formation but also their morphology. 

These results are consistent with an increased elliptical-to-spiral ratio within filaments compared to the field, a result that is well established in the literature \citep{kuutma_voids_2017,ricciardelli_morphological_2017}.The simulation work of \cite{song_beyond_2021} suggests that galaxies in the center of filaments tend to have a more compact stellar distribution, which could be the result of efficient angular-momentum cancellation from filamentary flows.  All these results suggest that the filament environment does play a role in shaping the morphology of galaxies. This, however, is not clearly seen in the work of \citet{alpaslan_galaxy_2015}, which finds that, after normalisation in mass, the morphologies of galaxies in non-group environments (e.g., filaments and voids) are largely similar. However, these authors do find small differences in ellipticity, possibly suggesting that discs may be more likely to be found in voids, in some agreement with our findings. 

We find that galaxies in cluster interiors tend to possess the highest fraction of early-type galaxies. Given the local galaxy density distributions shown in \autoref{fig:Density_Distributions}, this is not surprising, as highlighted by the Morphology--Density relation that galaxies in more dense environments favour early-type morphologies. To determine whether or not these differences, as well as those seen in filaments, are a consequence of the processes associated with the large-scale or small-scale environment, we must account for the differences in local galaxy density distributions, as we did before. To accomplish this, we construct samples matched both in stellar mass and local galaxy density.

\subsubsection{T-Type -- mass and local-density matched samples}

\begin{figure*}
	\includegraphics[width=1.8\columnwidth, keepaspectratio]{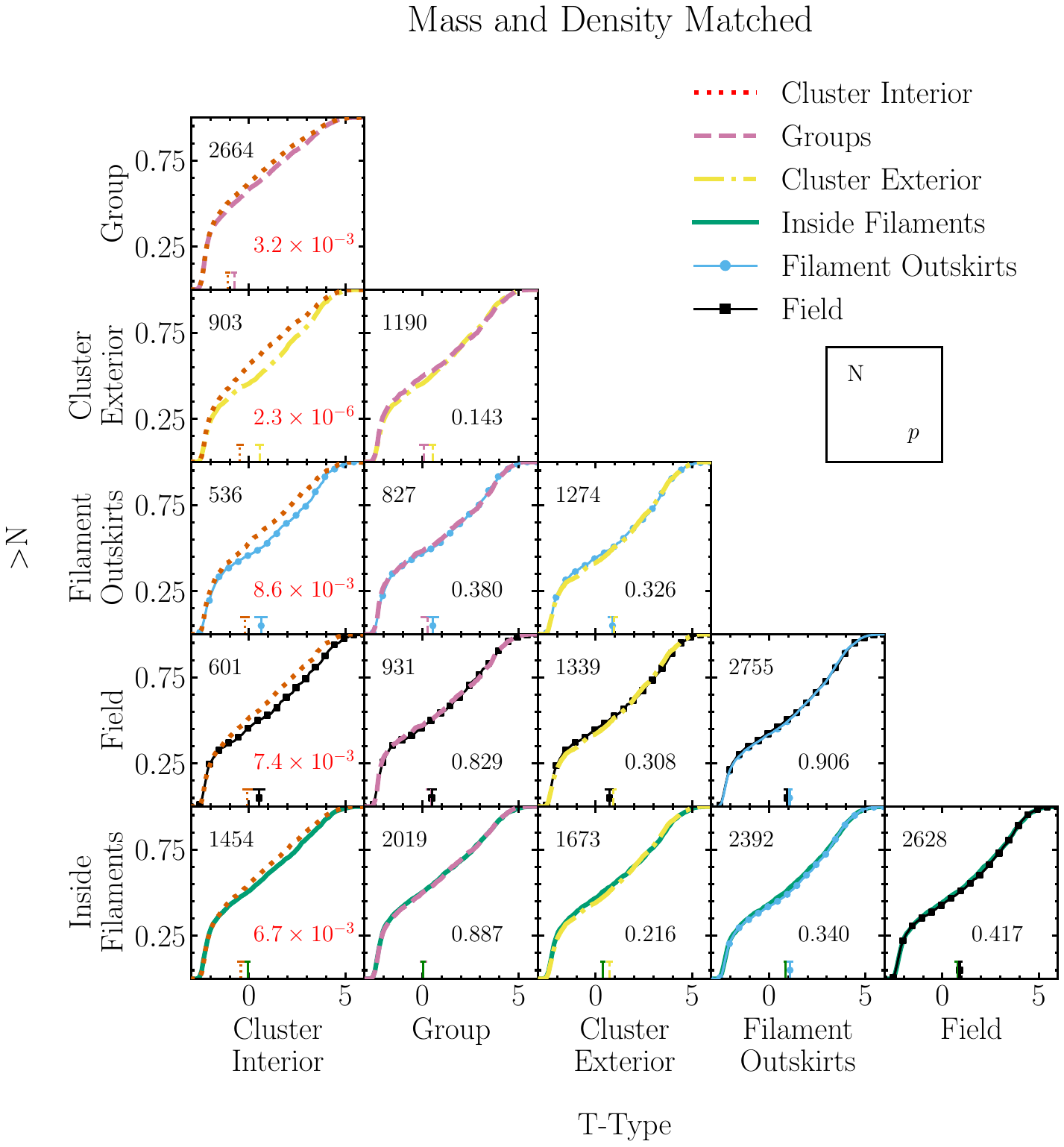}
    \caption{Pairwise comparisons of the T-Type distributions for each environmental bin pair, using the mass and local-density matched samples. Medians for each distribution are shown as vertical lines along with their respective $1\sigma$ errors. In each panel, the number of galaxies in each population is shown in the top left. Using Kolmogorov-Smirnov statistics, the probability that both distributions are identical is shown in the lower right. Significant p-values ($p < 0.05$) are coloured in blue and highly significant p-values ($p < 0.01$) are coloured in red. When matching in local galaxy density also, we find that the difference in the T-Type distributions between the filament and field population seen in the mass matched comparison (\autoref{fig:PairwiseTType}), vanishes. This suggests that the environmental effect of filaments on galaxy T-Type can be entirely parameterised by a local galaxy density index $\Sigma^*_3$.}
    \label{fig:PairwiseTType}
\end{figure*}

We find that the effects of filaments can be entirely encoded within a local galaxy density index. This is evident in \autoref{fig:PairwiseTType}, where we repeat the above analysis for samples matched both in stellar mass and local galaxy density. We find that the differences observed between the filament and field populations in the mass-matched samples vanishes when also matching in local galaxy density. This suggests that similarly to $\Delta$SFMS, the differences observed in the mass-matched samples are a consequence of the different local galaxy density distributions between the filament and field sample, as is expected given the Morphology--Density relation.

This further supports the conclusions discussed in section \ref{subsubsection: SFMS}, in which the environmental effect of filaments can entirely be characterised by local galaxy density. The interpretation and the discussion presented there of the possible physical mechanisms at play is also valid here. 

Similarly to what we found for $\Delta$SFMS, the only differences between the T-Type cumulative distributions can be found when comparing cluster interiors with lower-density environments. This suggests that galaxies in the clusters are subject to additional environmental effects affecting their morphology that are not characterised by local galaxy density alone. Gravitational tidal effects due to the cluster potential is a plausible mechanism for this, in addition to ram-pressure stripping. 

Numerous past studies have attempted to account for the effect of local galaxy density in this context; one such is the work of \citet{kuutma_voids_2017}, which finds that the Elliptical-to-Spiral ratio decreases with increasing distance from filaments, after normalising in both mass and density. We note that this is not necessarily in disagreement with our work, as we opt to identify galaxies within filaments which are also members of groups or clusters and consider them separately, and not as members of the filaments themselves. As such, the trends observed by \citet{kuutma_voids_2017} may be driven by the group and cluster populations within filaments. We further note that this discrepancy may be due to differences in methodologies also, such as different environmental density and filament sample definitions.

Additionally, \citet{Castignani_Virgo_2} investigate how galaxy morphology varies in filaments around the Virgo cluster as a function of local galaxy density. The authors conclude that filament galaxies tend to have a decreased late-type fraction compared to the field population. However, these results are at fixed local galaxy density only, without accounting for stellar mass. It has been shown in this work (Figure \ref{fig:Mass_Cumulative}) as well as previous works such as \citet{alpaslan_galaxy_2015} that stellar mass distributions tend to vary between cosmic web environments. It is possible then that the difference seen by \citet{Castignani_Virgo_2} could be a consequence of different stellar mass distributions, different methodologies, and/or the result of including lower mass galaxies in their sample given the proximity of the Virgo cluster. We discuss the implications of considering lower mass galaxies than those in the present work in section \ref{sec: pre-processing}.

As we did in section \ref{Sec: Results - SFMS}, we find that swapping to the local number density index $\Sigma_n$ does not affect the conclusions of this work. We do find small changes when going from $\Sigma^*_3$ to $\Sigma^*_5$, we find that the differences between cluster interiors and groups vanishes, while we observe only a small but statistically insignificant ($p = 0.272$) difference between cluster interiors and the field. It is likely the latter is a result of the small overlap in density distributions at this larger scale. The lack of difference between cluster interiors and groups at the larger scale could be a result of the inability to find analogues for the densest, most central of galaxies in groups at this larger scale. This hints that the difference seen between cluster interiors and groups could be driven by galaxies residing in the densest, most central regions of clusters. However, we stress that the main objective is investigating the effects of filaments, the conclusions of which are unchanged with the different local galaxy density indices tested above.

\subsection{Pre-processing and the effect of filaments} \label{sec: pre-processing}

We consider now what these results mean for pre-processing within filaments. We must emphasise that observational studies, such as this work, are restricted to the present environment only, whilst this does indeed correlate with the historic environment, it does not fully describe it. We must therefore be cautious in allocating the effects of the current and observed environment, which actually may be a result of the past environment. 

In Figures~\ref{fig:SFMS_Cumulative} and~\ref{fig:TTYPE_Cumulative} we find that galaxies in filaments tend to be suppressed in star formation and favour earlier-type morphologies relative to the field population. We further show that this effect persists when matching in stellar mass in Figures~\ref{fig:MassOnlyPairwiseSFMS} and~\ref{fig:MassOnlyPairwiseTType}. This suggests that galaxies in filaments could be subject to environmental effects and that filaments of the cosmic web serve as important intermediate environments for galaxy evolution. 

We find however in Figures~\ref{fig:PairwiseSFMS} and~\ref{fig:PairwiseTType} that when matching also in local galaxy density, the differences between the filament and field populations vanish. We conclude that the effects of filaments can be entirely encoded within a local galaxy density index, suggesting that the effects within the mass-matched sample are a consequence of the Star Formation--Density and Morphology--Density relations due to the differing local galaxy density distributions. We find no evidence suggesting that filaments impart unique environmental effects that cannot be characterised by local galaxy density. Whilst this conclusion holds for the galaxy sample considered in this work, we note that our conclusions on the role of filaments are subject to our mass limit ($M_{\text{stellar}} > 10^{9.91} \text{M}_{\sun}$). Given that low mass galaxies are affected more strongly by mechanisms present in higher density regions of the cosmic web, this is important to consider. 

The efficiency of many environmental processes is dependant on the mass of the galaxy on which they act. One such example is ram-pressure stripping; high mass galaxies can better retain their gas due to the deeper potential wells \citep{fillingham_taking_2015}. Furthermore, we refer to works such as that of \citet{donnari_quenched_2021} and \citet{hasan_filaments_2023}, which investigate how galaxy properties vary with the environment using the IllustrisTNG hydrodynamical simulations, for insights from the theoretical perspective. \citeauthor{donnari_quenched_2021} find that at low redshifts, 30\% of quenched group and cluster satellites were already quenched before infalling onto their current host, concluding that this is due to pre-processing for low-mass galaxies (${M}_{\text{stellar}} \leq 10^{10.0-10.5} \text{M}_{\sun}$), whereas high-mass galaxies tend to quench independently of the environment or via AGN feedback. \citeauthor{hasan_filaments_2023} also find that quenching in high-mass galaxies is driven by mass, while lower-mass galaxies are more likely to be quenched in regions of higher filament linear density (i.e., thick filaments). A similar result is found by \citet{goubert_role_2024}, suggesting that intrinsic parameters such as black hole mass are the dominant predictor of quiescence in centrals and high-mass satellites, whereas quiescence in low-mass satellites is correlated with environmental parameters, a trend shown both in observations and simulations. This highlights the importance of considering the behaviour of both high-mass and low-mass galaxies when ascertaining the role of filaments in shaping galaxy properties.

The conclusions of this work are subject to the accuracy of our filament networks, we verify this following the reasoning of \citet{chen_detecting_2017}. Filament galaxies are expected to exhibit some distinct properties. For instance, filament galaxies are expected to be more cluster-like than those in the field. A valid filament network should reflect this. To test this, we construct test networks using a completely random distribution of tracers. We find that the results presented in this work comparing filament and field galaxies vanish when using this random test network. We therefore conclude that the filament networks in this work are indeed representative of physical structure. 

In this work, we are restricted to relatively high-mass galaxies. The mass limit we adopt is motivated by observational restraints. At the redshifts of the regions we study ($z \approx 0.05$), current wide-field spectroscopic surveys such as SDSS cannot provide reliable spectra for fainter low-mass galaxies. A question that we cannot answer yet is: could lower-mass galaxies show additional signatures of pre-processing not characterised by local density? A question we intend to answer through the observations collected through the WEAVE Wide-Field Cluster Survey.

\section{Conclusions and future work} \label{sec: Conclusions}

In this work, we have investigated how the properties of 23,441 galaxies in the SDSS DR8 Main Galaxy Sample vary as a function of their environment. The stellar mass limit of our sample galaxies is $M_{\text{stellar}} > 10^{9.91} \text{M}_{\sun}$ and galaxies are selected in a narrow redshift slice about 6 of the WEAVE Wide-Field Cluster Survey target clusters ($z\sim0.05$). We used \texttt{DisPerSE} to extract the 2D cosmic web in regions within $\sim 100\times100\,$Mpc$^2$ area around the target clusters. 

We show in \autoref{fig:SFMS_Cumulative} that galaxies inside filaments tend to be more suppressed in star formation relative to those in the field, and enhanced relative to those in groups and clusters. We find parallel trends in galaxy morphology (T-Type; \autoref{fig:TTYPE_Cumulative}): galaxies in filaments favour earlier-type morphologies relative to the field population.

We also find that stellar mass distributions of galaxies within each environment differ, with galaxies in filaments tending to be less massive than those in cluster interiors and groups. We also find a hint that filament galaxies tend to be slightly more massive than those in the field, this difference however is not statistically significant. We show that when accounting for the differences in stellar mass distributions through constructing mass-matched samples, the differences in $\Delta$SFMS (measuring star-formation suppression) and T-Type persist. This is presented in Figures~\ref{fig:MassOnlyPairwiseSFMS} and~\ref{fig:MassOnlyPairwiseTType}, showing that galaxies in filaments differ from those in the field and those in groups and clusters, even at fixed stellar mass. This result agrees with that of numerous past studies and suggests that galaxies in filaments are subject to pre-processing. 

We have investigated whether these differences are a consequence of the well-established SFR--Density and Morphology--Density relations by constructing mass- and local-density-matched galaxy samples. Whilst previous studies compute local densities on large scales ($>\,$Mpc), we compute densities on smaller scales ($\leq\,$Mpc), to probe the most recent, local, and stochastic physical processes. These results are presented in Figures~\ref{fig:PairwiseSFMS} and~\ref{fig:PairwiseTType}. We show that the differences between properties of galaxies in filaments and those in the field found in the mass-matched samples vanish when also matching in local galaxy density. This indicates that the effect of the filament environment can be entirely encoded within a local galaxy density index. 

We find that in the mass- and density-matched samples, significant differences can only be seen when comparing cluster interiors with lower-density environments. We thus conclude that the environmental effects on both star formation and morphology can be entirely characterised by local galaxy density except in the interiors of clusters. This suggests that these are unique environments, with additional physical processes such as ram-pressure stripping or strong tidal effects, which do not act significantly in other environments. 

At this stage, however, we cannot make firm conclusions as to what physical processes are responsible for the results observed in this work. We think that our discussion of the possible physics is reasonable and plausible, given the available information, but speculative. However, there is a clear way forward to make progress in the future. In this work we have focused on present-day galaxy properties such as current star-formation rates and morphologies, and current environments. A natural extension would be to include temporal information through the analysis of the star-formation and chemical histories of the galaxies. This is possible with the help of physically motivated galaxy evolution models, constrained with extensive photometric and spectroscopic data,  such as the one presented by \citet{zhou_impact_2022}. We expect to be able to investigate and compare both star formation histories and timescales  across environments, allowing us to make firmer inferences about the physical processes at play.

The work presented here has given us very interesting insights on the effect of cosmic-web filaments on the transformation of relatively massive galaxies over large spatial scales. It also serves as a precursor of the WEAVE Wide-Field Cluster Survey, where we will explore the filament and group environment in the vicinity of low-redshift clusters. The WWFCS will prove invaluable in several ways. It will, first, provide a very large and robust statistical sample containing tens of thousands of lower-mass galaxies around clusters, reaching $M_{\text{stellar}}\sim10^{9} \text{M}_{\sun}$ up to $\sim5R_{200}$ from the cluster centres.  Its very high sampling density will allow us to map the cosmic web around clusters with exquisite detail. Its higher signal-to-noise spectra will yield detailed information on the star-formation and stellar-population properties of the galaxies, extending our current environmental research to galaxies with one order-of-magnitude smaller masses.

\section*{Acknowledgements}

We thank the anonymous reviewer for their helpful comments which greatly improved the quality of this paper. The authors are indebted to Phil Parry for the technical support with the installation of DisPerSE and Daniel Cornwell whose expertise with DisPerSE helped this work immeasurably. The authors further thank Prof. Frazer Pearce whose thought-provoking comments helped to improve the quality of this paper.

This work was supported by the Science and Technology Facilities Research Council [grant number: ST/X508639/1]. AAS, MG and UK acknowledge financial support from the UK Science and Technology Facilities Council (STFC; grant ref: ST/T000171/1)

Funding for SDSS-III has been provided by the Alfred P. Sloan Foundation, the Participating Institutions, the National Science Foundation, and the U.S. Department of Energy Office of Science. The SDSS-III web site is http://www.sdss3.org/.

SDSS-III is managed by the Astrophysical Research Consortium for the Participating Institutions of the SDSS-III Collaboration including the University of Arizona, the Brazilian Participation Group, Brookhaven National Laboratory, Carnegie Mellon University, University of Florida, the French Participation Group, the German Participation Group, Harvard University, the Instituto de Astrofisica de Canarias, the Michigan State/Notre Dame/JINA Participation Group, Johns Hopkins University, Lawrence Berkeley National Laboratory, Max Planck Institute for Astrophysics, Max Planck Institute for Extraterrestrial Physics, New Mexico State University, New York University, Ohio State University, Pennsylvania State University, University of Portsmouth, Princeton University, the Spanish Participation Group, University of Tokyo, University of Utah, Vanderbilt University, University of Virginia, University of Washington, and Yale University.
For the purpose of open access, the authors have applied a Creative Commons attribution (CC BY) licence to any Author Accepted Manuscript version arising. The authors contributed to this paper in the following ways: CJO, UK, AAS and MEG formed the core team. CJO analysed the data, produced the plots, and wrote the paper along with UK, AAS and MEG.

%%%%%%%%%%%%%%%%%%%%%%%%%%%%%%%%%%%%%%%%%%%%%%%%%%
\section*{Data Availability}

The data underlying this article were accessed from: SDSS DR8 http://www.sdss3.org/dr8/. The derived data generated in this research will be shared on request to the corresponding author.

%%%%%%%%%%%%%%%%%%%% REFERENCES %%%%%%%%%%%%%%%%%%

% The best way to enter references is to use BibTeX:

\bibliographystyle{mnras}
\bibliography{references} % if your bibtex file is called example.bib

\begin{thebibliography}{}
\makeatletter
\relax
\def\mn@urlcharsother{\let\do\@makeother \do\$\do\&\do\#\do\^\do\_\do\%\do\~}
\def\mn@doi{\begingroup\mn@urlcharsother \@ifnextchar [ {\mn@doi@} {\mn@doi@[]}}
\def\mn@doi@[#1]#2{\def\@tempa{#1}\ifx\@tempa\@empty \href {http://dx.doi.org/#2} {doi:#2}\else \href {http://dx.doi.org/#2} {#1}\fi \endgroup}
\def\mn@eprint#1#2{\mn@eprint@#1:#2::\@nil}
\def\mn@eprint@arXiv#1{\href {http://arxiv.org/abs/#1} {{\tt arXiv:#1}}}
\def\mn@eprint@dblp#1{\href {http://dblp.uni-trier.de/rec/bibtex/#1.xml} {dblp:#1}}
\def\mn@eprint@#1:#2:#3:#4\@nil{\def\@tempa {#1}\def\@tempb {#2}\def\@tempc {#3}\ifx \@tempc \@empty \let \@tempc \@tempb \let \@tempb \@tempa \fi \ifx \@tempb \@empty \def\@tempb {arXiv}\fi \@ifundefined {mn@eprint@\@tempb}{\@tempb:\@tempc}{\expandafter \expandafter \csname mn@eprint@\@tempb\endcsname \expandafter{\@tempc}}}

\bibitem[\protect\citeauthoryear{Aihara et~al.,}{Aihara et~al.}{2011}]{aihara_eighth_2011}
Aihara H.,  et~al., 2011, \mn@doi [The Astrophysical Journal Supplement Series] {10.1088/0067-0049/193/2/29}, 193, 29

\bibitem[\protect\citeauthoryear{Alberts \& Noble}{Alberts \& Noble}{2022}]{alberts_clusters_2022}
Alberts S.,  Noble A.,  2022, \mn@doi [Universe] {10.3390/universe8110554}, 8, 554

\bibitem[\protect\citeauthoryear{Alpaslan et~al.,}{Alpaslan et~al.}{2015}]{alpaslan_galaxy_2015}
Alpaslan M.,  et~al., 2015, \mn@doi [Monthly Notices of the Royal Astronomical Society] {10.1093/mnras/stv1176}, 451, 3249

\bibitem[\protect\citeauthoryear{Aragón-Calvo, van~de Weygaert  \& Jones}{Aragón-Calvo et~al.}{2010}]{aragon-calvo_multiscale_2010}
Aragón-Calvo M.~A.,  van~de Weygaert R.,   Jones B. J.~T.,  2010, \mn@doi [Monthly Notices of the Royal Astronomical Society] {10.1111/j.1365-2966.2010.17263.x}, 408, 2163

\bibitem[\protect\citeauthoryear{Bahe, McCarthy, Balogh  \& Font}{Bahe et~al.}{2013}]{bahe_why_2013}
Bahe Y.~M.,  McCarthy I.~G.,  Balogh M.~L.,   Font A.~S.,  2013, \mn@doi [MONTHLY NOTICES OF THE ROYAL ASTRONOMICAL SOCIETY] {10.1093/mnras/stt109}, 430, 3017

\bibitem[\protect\citeauthoryear{Baldry, Balogh, Bower, Glazebrook, Nichol, Bamford  \& Budavari}{Baldry et~al.}{2006}]{baldry_galaxy_2006}
Baldry I.~K.,  Balogh M.~L.,  Bower R.~G.,  Glazebrook K.,  Nichol R.~C.,  Bamford S.~P.,   Budavari T.,  2006, \mn@doi [Monthly Notices of the Royal Astronomical Society] {10.1111/j.1365-2966.2006.11081.x}, 373, 469

\bibitem[\protect\citeauthoryear{Bamford et~al.,}{Bamford et~al.}{2009}]{bamford_galaxy_2009}
Bamford S.~P.,  et~al., 2009, \mn@doi [Monthly Notices of the Royal Astronomical Society] {10.1111/j.1365-2966.2008.14252.x}, 393, 1324

\bibitem[\protect\citeauthoryear{Barsanti et~al.,}{Barsanti et~al.}{2023}]{barsanti_sami_2023}
Barsanti S.,  et~al., 2023, The {SAMI} {Galaxy} {Survey}: impact of black hole activity on galaxy spin-filament alignments, \url {http://arxiv.org/abs/2309.02794}

\bibitem[\protect\citeauthoryear{Blanton et~al.,}{Blanton et~al.}{2005}]{blanton_new_2005}
Blanton M.~R.,  et~al., 2005, \mn@doi [The Astronomical Journal] {10.1086/429803}, 129, 2562

\bibitem[\protect\citeauthoryear{Bond, Kofman  \& Pogosyan}{Bond et~al.}{1996}]{bond_how_1996}
Bond J.~R.,  Kofman L.,   Pogosyan D.,  1996, \mn@doi [Nature] {10.1038/380603a0}, 380, 603

\bibitem[\protect\citeauthoryear{Bond, Strauss  \& Cen}{Bond et~al.}{2010}]{bond_crawling_2010}
Bond N.~A.,  Strauss M.~A.,   Cen R.,  2010, \mn@doi [Monthly Notices of the Royal Astronomical Society] {10.1111/j.1365-2966.2010.17307.x}, 409, 156

\bibitem[\protect\citeauthoryear{Brinchmann, Charlot, White, Tremonti, Kauffmann, Heckman  \& Brinkmann}{Brinchmann et~al.}{2004}]{brinchmann_physical_2004}
Brinchmann J.,  Charlot S.,  White S. D.~M.,  Tremonti C.,  Kauffmann G.,  Heckman T.,   Brinkmann J.,  2004, \mn@doi [Monthly Notices of the Royal Astronomical Society] {10.1111/j.1365-2966.2004.07881.x}, 351, 1151

\bibitem[\protect\citeauthoryear{Brown et~al.,}{Brown et~al.}{2023}]{brown_vertico_2023}
Brown T.,  et~al., 2023, {VERTICO} {VII}: {Environmental} quenching caused by suppression of molecular gas content and star formation efficiency in {Virgo} {Cluster} galaxies, \url {http://arxiv.org/abs/2308.10943}

\bibitem[\protect\citeauthoryear{Bulichi, Dave  \& Kraljic}{Bulichi et~al.}{2023}]{bulichi_how_2023}
Bulichi T.-E.,  Dave R.,   Kraljic K.,  2023, How galaxy properties vary with filament proximity in the {SIMBA} simulations, \url {http://arxiv.org/abs/2309.03282}

\bibitem[\protect\citeauthoryear{Castignani et~al.,}{Castignani et~al.}{2022a}]{Castignani_Virgo_2}
Castignani G.,  et~al., 2022a, \mn@doi [The Astrophysical Journal Supplement Series] {10.3847/1538-4365/ac45f7}, 259, 43

\bibitem[\protect\citeauthoryear{Castignani et~al.,}{Castignani et~al.}{2022b}]{Castignani_Virgo_1}
Castignani G.,  et~al., 2022b, \mn@doi [Astronomy \& Astrophysics] {10.1051/0004-6361/202040141}, 657, A9

\bibitem[\protect\citeauthoryear{Cautun, van~de Weygaert, Jones  \& Frenk}{Cautun et~al.}{2014}]{cautun_evolution_2014}
Cautun M.,  van~de Weygaert R.,  Jones B. J.~T.,   Frenk C.~S.,  2014, \mn@doi [Monthly Notices of the Royal Astronomical Society] {10.1093/mnras/stu768}, 441, 2923

\bibitem[\protect\citeauthoryear{Chen et~al.,}{Chen et~al.}{2017}]{chen_detecting_2017}
Chen Y.-C.,  et~al., 2017, \mn@doi [Monthly Notices of the Royal Astronomical Society] {10.1093/mnras/stw3127}, 466, 1880

\bibitem[\protect\citeauthoryear{Chung, Kim, Rey  \& Lee}{Chung et~al.}{2021}]{chung_star-forming_2021}
Chung J.,  Kim S.,  Rey S.-C.,   Lee Y.,  2021, \mn@doi [The Astrophysical Journal] {10.3847/1538-4357/ac3002}, 923, 235

\bibitem[\protect\citeauthoryear{Colberg, Krughoff  \& Connolly}{Colberg et~al.}{2005}]{colberg_inter-cluster_2005}
Colberg J.~M.,  Krughoff K.~S.,   Connolly A.~J.,  2005, \mn@doi [Monthly Notices of the Royal Astronomical Society] {10.1111/j.1365-2966.2005.08897.x}, 359, 272

\bibitem[\protect\citeauthoryear{Colless et~al.,}{Colless et~al.}{2001}]{colless_2df_2001}
Colless M.,  et~al., 2001, \mn@doi [Monthly Notices of the Royal Astronomical Society] {10.1046/j.1365-8711.2001.04902.x}, 328, 1039

\bibitem[\protect\citeauthoryear{Cornwell et~al.,}{Cornwell et~al.}{2022}]{cornwell_forecasting_2022}
Cornwell D.~J.,  et~al., 2022, Forecasting the success of the {WEAVE} {Wide}-{Field} {Cluster} {Survey} on the extraction of the cosmic web filaments around galaxy clusters, \url {http://arxiv.org/abs/2209.13473}

\bibitem[\protect\citeauthoryear{Croton et~al.,}{Croton et~al.}{2006}]{croton_many_2006}
Croton D.~J.,  et~al., 2006, \mn@doi [Monthly Notices of the Royal Astronomical Society] {10.1111/j.1365-2966.2005.09675.x}, 365, 11

\bibitem[\protect\citeauthoryear{Dalton et~al.,}{Dalton et~al.}{2014}]{Dalton_WEAVE}
Dalton G.,  et~al., 2014. Montréal, Quebec, Canada, p. 91470L, \mn@doi{10.1117/12.2055132}, \url {http://proceedings.spiedigitallibrary.org/proceeding.aspx?doi=10.1117/12.2055132}

\bibitem[\protect\citeauthoryear{Darvish, Sobral, Mobasher, Scoville, Best, Sales  \& Smail}{Darvish et~al.}{2014}]{darvish_cosmic_2014}
Darvish B.,  Sobral D.,  Mobasher B.,  Scoville N.~Z.,  Best P.,  Sales L.~V.,   Smail I.,  2014, \mn@doi [The Astrophysical Journal] {10.1088/0004-637X/796/1/51}, 796, 51

\bibitem[\protect\citeauthoryear{Darvish, Mobasher, Sobral, Rettura, Scoville, Faisst  \& Capak}{Darvish et~al.}{2016}]{darvish_effects_2016}
Darvish B.,  Mobasher B.,  Sobral D.,  Rettura A.,  Scoville N.,  Faisst A.,   Capak P.,  2016, \mn@doi [The Astrophysical Journal] {10.3847/0004-637X/825/2/113}, 825, 113

\bibitem[\protect\citeauthoryear{Domínguez~Sánchez, Huertas-Company, Bernardi, Tuccillo  \& Fischer}{Domínguez~Sánchez et~al.}{2018}]{dominguez_sanchez_improving_2018}
Domínguez~Sánchez H.,  Huertas-Company M.,  Bernardi M.,  Tuccillo D.,   Fischer J.~L.,  2018, \mn@doi [Monthly Notices of the Royal Astronomical Society] {10.1093/mnras/sty338}, 476, 3661

\bibitem[\protect\citeauthoryear{Donnan, Tojeiro  \& Kraljic}{Donnan et~al.}{2022}]{donnan_role_2022}
Donnan C.~T.,  Tojeiro R.,   Kraljic K.,  2022, \mn@doi [Nature Astronomy] {10.1038/s41550-022-01619-w}, 6, 599

\bibitem[\protect\citeauthoryear{Donnari et~al.,}{Donnari et~al.}{2021}]{donnari_quenched_2021}
Donnari M.,  et~al., 2021, \mn@doi [Monthly Notices of the Royal Astronomical Society] {10.1093/mnras/staa3006}, 500, 4004

\bibitem[\protect\citeauthoryear{Dressler}{Dressler}{1980}]{dressler_galaxy_1980}
Dressler A.,  1980, \mn@doi [The Astrophysical Journal] {10.1086/157753}, 236, 351

\bibitem[\protect\citeauthoryear{Dressler, Oemler, Poggianti, Gladders, Abramson  \& Vulcani}{Dressler et~al.}{2013}]{dressler_imacs_2013}
Dressler A.,  Oemler A.,  Poggianti B.~M.,  Gladders M.~D.,  Abramson L.,   Vulcani B.,  2013, \mn@doi [The Astrophysical Journal] {10.1088/0004-637X/770/1/62}, 770, 62

\bibitem[\protect\citeauthoryear{Driver et~al.,}{Driver et~al.}{2009}]{driver_gama_2009}
Driver S.~P.,  et~al., 2009, \mn@doi [Astronomy and Geophysics] {10.1111/j.1468-4004.2009.50512.x}, 50, 5.12

\bibitem[\protect\citeauthoryear{Eardley et~al.,}{Eardley et~al.}{2015}]{eardley_galaxy_2015}
Eardley E.,  et~al., 2015, \mn@doi [Monthly Notices of the Royal Astronomical Society] {10.1093/mnras/stv237}, 448, 3665

\bibitem[\protect\citeauthoryear{Fadda, Biviano, Marleau, Storrie-Lombardi  \& Durret}{Fadda et~al.}{2008}]{fadda_starburst_2008}
Fadda D.,  Biviano A.,  Marleau F.~R.,  Storrie-Lombardi L.~J.,   Durret F.,  2008, \mn@doi [The Astrophysical Journal] {10.1086/526457}, 672, L9

\bibitem[\protect\citeauthoryear{Fasano et~al.,}{Fasano et~al.}{2006}]{fasano_wings_2006}
Fasano G.,  et~al., 2006, \mn@doi [Astronomy \& Astrophysics] {10.1051/0004-6361:20053816}, 445, 805

\bibitem[\protect\citeauthoryear{Fillingham, Cooper, Wheeler, Garrison-Kimmel, Boylan-Kolchin  \& Bullock}{Fillingham et~al.}{2015}]{fillingham_taking_2015}
Fillingham S.~P.,  Cooper M.~C.,  Wheeler C.,  Garrison-Kimmel S.,  Boylan-Kolchin M.,   Bullock J.~S.,  2015, \mn@doi [Monthly Notices of the Royal Astronomical Society] {10.1093/mnras/stv2058}, 454, 2039

\bibitem[\protect\citeauthoryear{Fujita}{Fujita}{2004}]{fujita_pre-processing_2004}
Fujita Y.,  2004, \mn@doi [Publications of the Astronomical Society of Japan] {10.1093/pasj/56.1.29}, 56, 29

\bibitem[\protect\citeauthoryear{Galárraga-Espinosa et~al.,}{Galárraga-Espinosa et~al.}{2023a}]{galarraga-espinosa_evolution_2023}
Galárraga-Espinosa D.,  et~al., 2023a, Evolution of cosmic filaments in the {MTNG} simulation, \url {http://arxiv.org/abs/2309.08659}

\bibitem[\protect\citeauthoryear{Galárraga-Espinosa, Garaldi  \& Kauffmann}{Galárraga-Espinosa et~al.}{2023b}]{galarraga-espinosa_flows_2023}
Galárraga-Espinosa D.,  Garaldi E.,   Kauffmann G.,  2023b, \mn@doi [Astronomy \& Astrophysics] {10.1051/0004-6361/202244935}, 671, A160

\bibitem[\protect\citeauthoryear{Gill, Knebe  \& Gibson}{Gill et~al.}{2005}]{gill_evolution_2005}
Gill S. P.~D.,  Knebe A.,   Gibson B.~K.,  2005, \mn@doi [Monthly Notices of the Royal Astronomical Society] {10.1111/j.1365-2966.2004.08562.x}, 356, 1327

\bibitem[\protect\citeauthoryear{Gonzalez \& Padilla}{Gonzalez \& Padilla}{2010}]{gonzalez_automated_2010}
Gonzalez R.~E.,  Padilla N.~E.,  2010, \mn@doi [Monthly Notices of the Royal Astronomical Society] {10.1111/j.1365-2966.2010.17015.x}, 407, 1449

\bibitem[\protect\citeauthoryear{Goubert, Bluck, Piotrowska  \& Maiolino}{Goubert et~al.}{2024}]{goubert_role_2024}
Goubert P.~H.,  Bluck A. F.~L.,  Piotrowska J.~M.,   Maiolino R.,  2024, The role of environment and {AGN} feedback in quenching local galaxies: {Comparing} cosmological hydrodynamical simulations to the {SDSS}, \url {http://arxiv.org/abs/2401.12953}

\bibitem[\protect\citeauthoryear{Gunn \& Gott}{Gunn \& Gott}{1972}]{gunn_infall_1972}
Gunn J.~E.,  Gott Iii J.~R.,  1972, \mn@doi [The Astrophysical Journal] {10.1086/151605}, 176, 1

\bibitem[\protect\citeauthoryear{Haas, Schaye  \& Jeeson-Daniel}{Haas et~al.}{2012}]{haas_disentangling_2012}
Haas M.~R.,  Schaye J.,   Jeeson-Daniel A.,  2012, \mn@doi [Monthly Notices of the Royal Astronomical Society] {10.1111/j.1365-2966.2011.19863.x}, 419, 2133

\bibitem[\protect\citeauthoryear{Hasan et~al.,}{Hasan et~al.}{2023b}]{hasan_filaments_2023}
Hasan F.,  et~al., 2023b, Filaments of {The} {Slime} {Mold} {Cosmic} {Web} {And} {How} {They} {Affect} {Galaxy} {Evolution}, \url {http://arxiv.org/abs/2311.01443}

\bibitem[\protect\citeauthoryear{Hasan et~al.,}{Hasan et~al.}{2023a}]{hasan_how_2023}
Hasan F.,  et~al., 2023a, How {Cosmic} {Web} {Environment} {Affects} {Galaxy} {Quenching} {Across} {Cosmic} {Time}, \url {http://arxiv.org/abs/2303.08088}

\bibitem[\protect\citeauthoryear{Hashimoto, Augustus~Oemler, Lin  \& Tucker}{Hashimoto et~al.}{1998}]{hashimoto_influence_1998}
Hashimoto Y.,  Augustus~Oemler J.,  Lin H.,   Tucker D.~L.,  1998, \mn@doi [The Astrophysical Journal] {10.1086/305657}, 499, 589

\bibitem[\protect\citeauthoryear{Hinshaw et~al.,}{Hinshaw et~al.}{2012}]{hinshaw_nine-year_2012}
Hinshaw G.,  et~al., 2012, Nine-{Year} {Wilkinson} {Microwave} {Anisotropy} {Probe} ({WMAP}) {Observations}: {Cosmological} {Parameter} {Results}, \mn@doi{10.1088/0067-0049/208/2/19}, \url {https://arxiv.org/abs/1212.5226v3}

\bibitem[\protect\citeauthoryear{Hoosain et~al.,}{Hoosain et~al.}{2024}]{hoosain_effect_2024}
Hoosain M.,  et~al., 2024, The effect of cosmic web filaments on galaxy properties in the {RESOLVE} and {ECO} surveys, \url {http://arxiv.org/abs/2401.09114}

\bibitem[\protect\citeauthoryear{Jackson}{Jackson}{1972}]{jackson_fingers_1972}
Jackson J.~C.,  1972, \mn@doi [Monthly Notices of the Royal Astronomical Society] {10.1093/mnras/156.1.1P}, 156, 1P

\bibitem[\protect\citeauthoryear{Jin et~al.,}{Jin et~al.}{2023}]{jin_wide-field_2023}
Jin S.,  et~al., 2023, \mn@doi [Monthly Notices of the Royal Astronomical Society] {10.1093/mnras/stad557}, p. stad557

\bibitem[\protect\citeauthoryear{Kaiser}{Kaiser}{1984}]{kaiser_spatial_1984}
Kaiser N.,  1984, \mn@doi [The Astrophysical Journal] {10.1086/184341}, 284, L9

\bibitem[\protect\citeauthoryear{Kauffmann et~al.,}{Kauffmann et~al.}{2003}]{kauffmann_stellar_2003}
Kauffmann G.,  et~al., 2003, \mn@doi [Monthly Notices of the Royal Astronomical Society] {10.1046/j.1365-8711.2003.06291.x}, 341, 33

\bibitem[\protect\citeauthoryear{Kauffmann, White, Heckman, Ménard, Brinchmann, Charlot, Tremonti  \& Brinkmann}{Kauffmann et~al.}{2004}]{kauffmann_environmental_2004}
Kauffmann G.,  White S. D.~M.,  Heckman T.~M.,  Ménard B.,  Brinchmann J.,  Charlot S.,  Tremonti C.,   Brinkmann J.,  2004, \mn@doi [Monthly Notices of the Royal Astronomical Society] {10.1111/j.1365-2966.2004.08117.x}, 353, 713

\bibitem[\protect\citeauthoryear{Kim et~al.,}{Kim et~al.}{2016}]{kim_large-scale_2016}
Kim S.,  et~al., 2016, \mn@doi [The Astrophysical Journal] {10.3847/1538-4357/833/2/207}, 833, 207

\bibitem[\protect\citeauthoryear{Kleiner, Pimbblet, Jones, Koribalski  \& Serra}{Kleiner et~al.}{2017}]{kleiner_evidence_2017}
Kleiner D.,  Pimbblet K.~A.,  Jones D.~H.,  Koribalski B.~S.,   Serra P.,  2017, \mn@doi [Monthly Notices of the Royal Astronomical Society] {10.1093/mnras/stw3328}, 466, 4692

\bibitem[\protect\citeauthoryear{Kotecha et~al.,}{Kotecha et~al.}{2022}]{kotecha_cosmic_2022}
Kotecha S.,  et~al., 2022, \mn@doi [Monthly Notices of the Royal Astronomical Society] {10.1093/mnras/stac300}, 512, 926

\bibitem[\protect\citeauthoryear{Kraljic et~al.,}{Kraljic et~al.}{2018}]{kraljic_galaxy_2018}
Kraljic K.,  et~al., 2018, \mn@doi [Monthly Notices of the Royal Astronomical Society] {10.1093/mnras/stx2638}, 474, 547

\bibitem[\protect\citeauthoryear{Kuchner et~al.,}{Kuchner et~al.}{2020}]{kuchner_mapping_2020}
Kuchner U.,  et~al., 2020, \mn@doi [Monthly Notices of the Royal Astronomical Society] {10.1093/mnras/staa1083}, 494, 5473

\bibitem[\protect\citeauthoryear{Kuchner et~al.,}{Kuchner et~al.}{2021a}]{kuchner_cosmic_2021}
Kuchner U.,  et~al., 2021a, \mn@doi [Monthly Notices of the Royal Astronomical Society] {10.1093/mnras/stab567}, 503, 2065

\bibitem[\protect\citeauthoryear{Kuchner et~al.,}{Kuchner et~al.}{2021b}]{kuchner_inventory_2021}
Kuchner U.,  et~al., 2021b, \mn@doi [Monthly Notices of the Royal Astronomical Society] {10.1093/mnras/stab3419}, 510, 581

\bibitem[\protect\citeauthoryear{Kuutma, Tamm  \& Tempel}{Kuutma et~al.}{2017}]{kuutma_voids_2017}
Kuutma T.,  Tamm A.,   Tempel E.,  2017, \mn@doi [Astronomy \& Astrophysics] {10.1051/0004-6361/201730526}, 600, L6

\bibitem[\protect\citeauthoryear{Laigle et~al.,}{Laigle et~al.}{2018}]{laigle_cosmos2015_2018}
Laigle C.,  et~al., 2018, \mn@doi [Monthly Notices of the Royal Astronomical Society] {10.1093/mnras/stx3055}, 474, 5437

\bibitem[\protect\citeauthoryear{Larson}{Larson}{1974}]{larson_effects_1974}
Larson R.~B.,  1974, \mn@doi [Monthly Notices of the Royal Astronomical Society] {10.1093/mnras/169.2.229}, 169, 229

\bibitem[\protect\citeauthoryear{Larson, Tinsley  \& Caldwell}{Larson et~al.}{1980}]{larson_evolution_1980}
Larson R.~B.,  Tinsley B.~M.,   Caldwell C.~N.,  1980, \mn@doi [The Astrophysical Journal] {10.1086/157917}, 237, 692

\bibitem[\protect\citeauthoryear{Libeskind et~al.,}{Libeskind et~al.}{2018}]{libeskind_tracing_2018}
Libeskind N.~I.,  et~al., 2018, \mn@doi [Monthly Notices of the Royal Astronomical Society] {10.1093/mnras/stx1976}, 473, 1195

\bibitem[\protect\citeauthoryear{Lu, Mandelker, Oh, Dekel, Bosch, Springel, Nagai  \& van~de Voort}{Lu et~al.}{2023}]{lu_structure_2023}
Lu Y.~S.,  Mandelker N.,  Oh S.~P.,  Dekel A.,  Bosch F. C. v.~d.,  Springel V.,  Nagai D.,   van~de Voort F.,  2023, The {Structure} and {Dynamics} of {Massive} {High}-\$z\$ {Cosmic}-{Web} {Filaments}: {Three} {Radial} {Zones} in {Filament} {Cross}-{Sections}, \url {http://arxiv.org/abs/2306.03966}

\bibitem[\protect\citeauthoryear{Luber, Van~Gorkom, Hess, Pisano, Fernández  \& Momjian}{Luber et~al.}{2019}]{luber_large-scale_2019}
Luber N.,  Van~Gorkom J.~H.,  Hess K.~M.,  Pisano D.~J.,  Fernández X.,   Momjian E.,  2019, \mn@doi [The Astronomical Journal] {10.3847/1538-3881/ab1b6e}, 157, 254

\bibitem[\protect\citeauthoryear{Mahajan, Singh  \& Shobhana}{Mahajan et~al.}{2018}]{mahajan_ultraviolet_2018}
Mahajan S.,  Singh A.,   Shobhana D.,  2018, \mn@doi [Monthly Notices of the Royal Astronomical Society] {10.1093/mnras/sty1370}

\bibitem[\protect\citeauthoryear{Malavasi et~al.,}{Malavasi et~al.}{2017}]{malavasi_vimos_2017}
Malavasi N.,  et~al., 2017, \mn@doi [Monthly Notices of the Royal Astronomical Society] {10.1093/mnras/stw2864}, 465, 3817

\bibitem[\protect\citeauthoryear{Martínez, Muriel  \& Coenda}{Martínez et~al.}{2016}]{martinez_galaxies_2016}
Martínez H.~J.,  Muriel H.,   Coenda V.,  2016, \mn@doi [Monthly Notices of the Royal Astronomical Society] {10.1093/mnras/stv2295}, 455, 127

\bibitem[\protect\citeauthoryear{McGee, Balogh, Bower, Font  \& McCarthy}{McGee et~al.}{2009}]{mcgee_accretion_2009}
McGee S.~L.,  Balogh M.~L.,  Bower R.~G.,  Font A.~S.,   McCarthy I.~G.,  2009, \mn@doi [Monthly Notices of the Royal Astronomical Society] {10.1111/j.1365-2966.2009.15507.x}, 400, 937

\bibitem[\protect\citeauthoryear{Moore, Katz, Lake, Dressler  \& Oemler}{Moore et~al.}{1996}]{moore_galaxy_1996}
Moore B.,  Katz N.,  Lake G.,  Dressler A.,   Oemler A.,  1996, \mn@doi [Nature] {10.1038/379613a0}, 379, 613

\bibitem[\protect\citeauthoryear{Moretti et~al.,}{Moretti et~al.}{2017}]{moretti_omegawings_2017}
Moretti A.,  et~al., 2017, \mn@doi [Astronomy \& Astrophysics] {10.1051/0004-6361/201630030}, 599, A81

\bibitem[\protect\citeauthoryear{Muldrew et~al.,}{Muldrew et~al.}{2012}]{muldrew_measures_2012}
Muldrew S.~I.,  et~al., 2012, \mn@doi [Monthly Notices of the Royal Astronomical Society] {10.1111/j.1365-2966.2011.19922.x}, 419, 2670

\bibitem[\protect\citeauthoryear{Nair \& Abraham}{Nair \& Abraham}{2010}]{nair_catalog_2010}
Nair P.~B.,  Abraham R.~G.,  2010, \mn@doi [The Astrophysical Journal Supplement Series] {10.1088/0067-0049/186/2/427}, 186, 427

\bibitem[\protect\citeauthoryear{Nulsen}{Nulsen}{1982}]{nulsen_transport_1982}
Nulsen P. E.~J.,  1982, \mn@doi [Monthly Notices of the Royal Astronomical Society] {10.1093/mnras/198.4.1007}, 198, 1007

\bibitem[\protect\citeauthoryear{Oesch et~al.,}{Oesch et~al.}{2010}]{oesch_buildup_2010}
Oesch P.~A.,  et~al., 2010, \mn@doi [The Astrophysical Journal Letters] {10.1088/2041-8205/714/1/L47}, 714, L47

\bibitem[\protect\citeauthoryear{Oyarzun, Bundy, Westfall, Lacerna, Yan, Brownstein, Drory  \& Lane}{Oyarzun et~al.}{2023}]{oyarzun_sdss-iv_2023}
Oyarzun G.~A.,  Bundy K.,  Westfall K.~B.,  Lacerna I.,  Yan R.,  Brownstein J.~R.,  Drory N.,   Lane R.~R.,  2023, \mn@doi [The Astrophysical Journal] {10.3847/1538-4357/acbbca}, 947, 13

\bibitem[\protect\citeauthoryear{Parente et~al.,}{Parente et~al.}{2023}]{parente_star_2023}
Parente M.,  et~al., 2023, Star {Formation} and {Dust} in the {Cosmic} {Web}, \url {http://arxiv.org/abs/2312.10146}

\bibitem[\protect\citeauthoryear{Peng et~al.,}{Peng et~al.}{2010}]{peng_mass_2010}
Peng Y.-j.,  et~al., 2010, \mn@doi [The Astrophysical Journal] {10.1088/0004-637X/721/1/193}, 721, 193

\bibitem[\protect\citeauthoryear{Pimbblet, Drinkwater  \& Hawkrigg}{Pimbblet et~al.}{2004}]{pimbblet_intercluster_2004}
Pimbblet K.~A.,  Drinkwater M.~J.,   Hawkrigg M.~C.,  2004, \mn@doi [Monthly Notices of the Royal Astronomical Society] {10.1111/j.1365-2966.2004.08425.x}, 354, L61

\bibitem[\protect\citeauthoryear{Poggianti et~al.,}{Poggianti et~al.}{2017}]{poggianti_gasp_2017}
Poggianti B.~M.,  et~al., 2017, \mn@doi [The Astrophysical Journal] {10.3847/1538-4357/aa78ed}, 844, 48

\bibitem[\protect\citeauthoryear{Popesso et~al.,}{Popesso et~al.}{2011}]{popesso_effect_2011}
Popesso P.,  et~al., 2011, \mn@doi [Astronomy \& Astrophysics] {10.1051/0004-6361/201015672}, 532, A145

\bibitem[\protect\citeauthoryear{Ramsøy, Slyz, Devriendt, Laigle  \& Dubois}{Ramsøy et~al.}{2021}]{ramsoy_rivers_2021}
Ramsøy M.,  Slyz A.,  Devriendt J.,  Laigle C.,   Dubois Y.,  2021, \mn@doi [Monthly Notices of the Royal Astronomical Society] {10.1093/mnras/stab015}, 502, 351

\bibitem[\protect\citeauthoryear{Ricciardelli, Cava, Varela  \& Tamone}{Ricciardelli et~al.}{2017}]{ricciardelli_morphological_2017}
Ricciardelli E.,  Cava A.,  Varela J.,   Tamone A.,  2017, \mn@doi [The Astrophysical Journal Letters] {10.3847/2041-8213/aa84ad}, 846, L4

\bibitem[\protect\citeauthoryear{Salim et~al.,}{Salim et~al.}{2007}]{salim_uv_2007}
Salim S.,  et~al., 2007, \mn@doi [The Astrophysical Journal Supplement Series] {10.1086/519218}, 173, 267

\bibitem[\protect\citeauthoryear{Sampaio, De~Carvalho, Ferreras, Aragón-Salamanca  \& Parker}{Sampaio et~al.}{2022}]{sampaio_blue_2022}
Sampaio V.~M.,  De~Carvalho R.~R.,  Ferreras I.,  Aragón-Salamanca A.,   Parker L.~C.,  2022, Monthly Notices of the Royal Astronomical Society, 509, 567

\bibitem[\protect\citeauthoryear{Sarron, Adami, Durret  \& Laigle}{Sarron et~al.}{2019}]{sarron_pre-processing_2019}
Sarron F.,  Adami C.,  Durret F.,   Laigle C.,  2019, \mn@doi [Astronomy \& Astrophysics] {10.1051/0004-6361/201935394}, 632, A49

\bibitem[\protect\citeauthoryear{Scholz-Díaz, Martín-Navarro  \& Falcón-Barroso}{Scholz-Díaz et~al.}{2022}]{scholz-diaz_dark_2022}
Scholz-Díaz L.,  Martín-Navarro I.,   Falcón-Barroso J.,  2022, \mn@doi [Monthly Notices of the Royal Astronomical Society] {10.1093/mnras/stac361}, 511, 4900

\bibitem[\protect\citeauthoryear{Singh, Mahajan  \& Bagla}{Singh et~al.}{2020}]{singh_study_2020}
Singh A.,  Mahajan S.,   Bagla J.~S.,  2020, \mn@doi [Monthly Notices of the Royal Astronomical Society] {10.1093/mnras/staa1913}, 497, 2265

\bibitem[\protect\citeauthoryear{Snedden, Coughlin, Phillips, Mathews  \& Suh}{Snedden et~al.}{2016}]{snedden_star_2016}
Snedden A.,  Coughlin J.,  Phillips L.~A.,  Mathews G.,   Suh I.-S.,  2016, \mn@doi [Monthly Notices of the Royal Astronomical Society] {10.1093/mnras/stv2421}, 455, 2804

\bibitem[\protect\citeauthoryear{Sobral, Best, Smail, Geach, Cirasuolo, Garn  \& Dalton}{Sobral et~al.}{2011}]{sobral_dependence_2011}
Sobral D.,  Best P.~N.,  Smail I.,  Geach J.~E.,  Cirasuolo M.,  Garn T.,   Dalton G.~B.,  2011, \mn@doi [Monthly Notices of the Royal Astronomical Society] {10.1111/j.1365-2966.2010.17707.x}, 411, 675

\bibitem[\protect\citeauthoryear{Song et~al.,}{Song et~al.}{2021}]{song_beyond_2021}
Song H.,  et~al., 2021, \mn@doi [Monthly Notices of the Royal Astronomical Society] {10.1093/mnras/staa3981}, 501, 4635

\bibitem[\protect\citeauthoryear{Sousbie}{Sousbie}{2011}]{sousbie_persistent_2011}
Sousbie T.,  2011, \mn@doi [Monthly Notices of the Royal Astronomical Society] {10.1111/j.1365-2966.2011.18394.x}, 414, 350

\bibitem[\protect\citeauthoryear{Strauss et~al.,}{Strauss et~al.}{2002}]{strauss_spectroscopic_2002}
Strauss M.~A.,  et~al., 2002, \mn@doi [The Astronomical Journal] {10.1086/342343}, 124, 1810

\bibitem[\protect\citeauthoryear{Szpila, Davé, Rennehan, Cui  \& Hough}{Szpila et~al.}{2024}]{szpila_nature_2024}
Szpila J.,  Davé R.,  Rennehan D.,  Cui W.,   Hough R.,  2024, The {Nature} and {Evolution} of {Early} {Massive} {Quenched} {Galaxies} in the {Simba}-{C} {Simulation}, \url {http://arxiv.org/abs/2402.08729}

\bibitem[\protect\citeauthoryear{Thomas, Maraston, Bender  \& Mendes~de Oliveira}{Thomas et~al.}{2005}]{thomas_epochs_2005}
Thomas D.,  Maraston C.,  Bender R.,   Mendes~de Oliveira C.,  2005, \mn@doi [The Astrophysical Journal] {10.1086/426932}, 621, 673

\bibitem[\protect\citeauthoryear{Thompson, Smith  \& Kraljic}{Thompson et~al.}{2023}]{thompson_gas_2023}
Thompson B.~B.,  Smith R.,   Kraljic K.,  2023, \mn@doi [Monthly Notices of the Royal Astronomical Society] {10.1093/mnras/stac2963}, 518, 1361

\bibitem[\protect\citeauthoryear{Trussler, Maiolino, Maraston, Peng, Thomas, Goddard  \& Lian}{Trussler et~al.}{2020}]{trussler_both_2020}
Trussler J.,  Maiolino R.,  Maraston C.,  Peng Y.,  Thomas D.,  Goddard D.,   Lian J.,  2020, \mn@doi [Monthly Notices of the Royal Astronomical Society] {10.1093/mnras/stz3286}, 491, 5406

\bibitem[\protect\citeauthoryear{Vulcani et~al.,}{Vulcani et~al.}{2019}]{vulcani_gasp_2019}
Vulcani B.,  et~al., 2019, \mn@doi [Monthly Notices of the Royal Astronomical Society] {10.1093/mnras/stz1399}, 487, 2278

\bibitem[\protect\citeauthoryear{Vulcani et~al.,}{Vulcani et~al.}{2020}]{vulcani_gasp_2020}
Vulcani B.,  et~al., 2020, \mn@doi [The Astrophysical Journal] {10.3847/1538-4357/ab7bdd}, 892, 146

\bibitem[\protect\citeauthoryear{Vulcani, Poggianti, Smith, Moretti, Jaffe, Gullieuszik, Fritz  \& Bellhouse}{Vulcani et~al.}{2022}]{vulcani_relevance_2022}
Vulcani B.,  Poggianti B.~M.,  Smith R.,  Moretti A.,  Jaffe Y.,  Gullieuszik M.,  Fritz J.,   Bellhouse C.,  2022, \mn@doi [The Astrophysical Journal] {10.3847/1538-4357/ac4809}, 927, 91

\bibitem[\protect\citeauthoryear{Wang, Wang  \& Chen}{Wang et~al.}{2023}]{wang_environmental_2023}
Wang K.,  Wang X.,   Chen Y.,  2023, Environmental dependence of the mass-metallicity relation in cosmological hydrodynamical simulations, \url {http://arxiv.org/abs/2305.08161}

\bibitem[\protect\citeauthoryear{Wang et~al.,}{Wang et~al.}{2024}]{wang_boundary_2024}
Wang W.,  et~al., 2024, The boundary of cosmic filaments, \url {http://arxiv.org/abs/2402.11678}

\bibitem[\protect\citeauthoryear{Wolf et~al.,}{Wolf et~al.}{2009}]{wolf_stages_2009}
Wolf C.,  et~al., 2009, \mn@doi [Monthly Notices of the Royal Astronomical Society] {10.1111/j.1365-2966.2008.14204.x}, 393, 1302

\bibitem[\protect\citeauthoryear{Wu, Jespersen  \& Wechsler}{Wu et~al.}{2024}]{wu_how_2024}
Wu J.~F.,  Jespersen C.~K.,   Wechsler R.~H.,  2024, How the {Galaxy}-{Halo} {Connection} {Depends} on {Large}-{Scale} {Environment}, \mn@doi{10.48550/arXiv.2402.07995}, \url {http://arxiv.org/abs/2402.07995}

\bibitem[\protect\citeauthoryear{Yang, Mo, van~den Bosch, Pasquali, Li  \& Barden}{Yang et~al.}{2007}]{yang_galaxy_2007}
Yang X.,  Mo H.~J.,  van~den Bosch F.~C.,  Pasquali A.,  Li C.,   Barden M.,  2007, \mn@doi [The Astrophysical Journal] {10.1086/522027}, 671, 153

\bibitem[\protect\citeauthoryear{York et~al.,}{York et~al.}{2000}]{york_sloan_2000}
York D.~G.,  et~al., 2000, \mn@doi [The Astronomical Journal] {10.1086/301513}, 120, 1579

\bibitem[\protect\citeauthoryear{Zel'dovich}{Zel'dovich}{1970}]{zeldovich_gravitational_1970}
Zel'dovich Y.~B.,  1970, Astronomy and Astrophysics, 5, 84

\bibitem[\protect\citeauthoryear{Zhou, Merrifield, Aragón-Salamanca, Brownstein, Drory, Yan  \& Lane}{Zhou et~al.}{2022}]{zhou_impact_2022}
Zhou S.,  Merrifield M.,  Aragón-Salamanca A.,  Brownstein J.~R.,  Drory N.,  Yan R.,   Lane R.~R.,  2022, The impact of environment on the lives of disk galaxies as revealed by {SDSS}-{IV} {MaNGA}, \url {http://arxiv.org/abs/2210.03509}

\bibitem[\protect\citeauthoryear{Zhou, Aragón-Salamanca  \& Merrifield}{Zhou et~al.}{2024}]{zhou_separate_2024}
Zhou S.,  Aragón-Salamanca A.,   Merrifield M.,  2024, The separate effect of halo mass and stellar mass on the evolution of massive disk galaxies, \mn@doi{10.48550/arXiv.2404.16181}, \url {http://arxiv.org/abs/2404.16181}

\bibitem[\protect\citeauthoryear{de Vaucouleurs}{de~Vaucouleurs}{1963}]{de_vaucouleurs_revised_1963}
de Vaucouleurs G.,  1963, \mn@doi [The Astrophysical Journal Supplement Series] {10.1086/190084}, 8, 31

\makeatother
\end{thebibliography}

% Alternatively you could enter them by hand, like this:
% This method is tedious and prone to error if you have lots of references
%\begin{thebibliography}{99}
%\bibitem[\protect\citeauthoryear{Author}{2012}]{Author2012}
%Author A.~N., 2013, Journal of Improbable Astronomy, 1, 1
%\bibitem[\protect\citeauthoryear{Others}{2013}]{Others2013}
%Others S., 2012, Journal of Interesting Stuff, 17, 198
%\end{thebibliography}

%%%%%%%%%%%%%%%%%%%%%%%%%%%%%%%%%%%%%%%%%%%%%%%%%%

%%%%%%%%%%%%%%%%% APPENDICES %%%%%%%%%%%%%%%%%%%%%

%\appendix

%%%%%%%%%%%%%%%%%%%%%%%%%%%%%%%%%%%%%%%%%%%%%%%%%%

% Don't change these lines
\bsp	% typesetting comment
\label{lastpage}
\end{document}